\documentclass[journal]{IEEEtran}

\usepackage{multicol}
\usepackage{indentfirst}
\usepackage[pdftex]{graphicx}
\usepackage{amsthm}
\usepackage{amsmath}
\usepackage{amssymb}
\usepackage{amsfonts}
\usepackage{mathrsfs}
\usepackage{graphicx}
\usepackage{enumerate}
\usepackage{multirow}
\usepackage{booktabs}
\usepackage{enumitem}
\usepackage{enumerate}
\usepackage[noend]{algpseudocode}
\usepackage{algorithmicx,algorithm}
\usepackage{bm}

\newtheorem{Theorem}{Theorem}

\newtheorem{Corollary}{Corollary}
\newtheorem*{Proof}{Proof}
\newtheorem*{Remark}{Remark}
\newtheorem*{Remarks}{Remarks}
\newtheorem*{Square Root Law}{Square Root Law}
\newtheorem{Lemma}{Lemma}
\newtheorem*{Triangular Inequality Bound}{Triangular Inequaliity Bound}
\newtheorem*{Data Processing Inequality Bound}{Data Processing Inequality Bound}

\ifCLASSINFOpdf
\else
\fi

\begin{document}
%
\title{Finite Blocklength Analysis of Gaussian Random Coding in AWGN Channels under Covert Constraint}
%
%
%

\author{Xinchun~Yu,
  Shuangqing Wei and Yuan~Luo  
        \thanks{This work was supported by National Natural Science Foundation of China under Grant 61871264.}
\thanks{Xinchun Yu and Yuan Luo are with the School of Electronic Information and Electrical Engineering, Shanghai Jiao Tong University, Shanghai 200240, China. Luo is the corresponding author. (e-mail: moonyuyu@sjtu.edu.cn; yuanluo@sjtu.edu.cn).

 Shuangqing Wei is with the Division of Electrical and Computer Engineering
School of Electrical Engineering and Computer Science,
Louisiana State University, Baton Rouge, LA 70803, USA (e-mail: swei@lsu.edu).}
}

\maketitle



%
\IEEEpeerreviewmaketitle

%
%
%
%

\begin{abstract}
It is well known that finite blocklength analysis plays an important role in evaluating performances of communication systems in practical settings. This paper considers the achievability and converse bounds on the maximal channel coding rate (throughput) at a given blocklength and error probability in covert communication over AWGN channels. The covert constraint is given in terms of an upper bound on total variation distance (TVD) between the distributions of eavesdropped signals at an adversary with and without presence of active and legitimate communication, respectively. For the achievability, Gaussian random coding scheme is adopted for convenience in the analysis of TVD. The classical results of finite blocklength regime are not applicable in this case. By exploiting and extending canonical approaches, we first present new and more general achievability bounds for random coding schemes under maximal or average probability of error requirements. The general bounds are then applied to covert communication in AWGN channels where codewords are generated from Gaussian distribution while meeting the maximal power constraint. We further show an interesting connection between attaining  tight achievability and converse bounds and solving two total variation distance based minimax and maxmin problems. The TVD constraint is analyzed under the given random coding scheme, which induces bounds on the transmission power through divergence inequalities. Further comparison is made between the new achievability bounds and existing ones derived under deterministic codebooks. Our thorough analysis thus leads us to a comprehensive characterization of the attainable throughput in covert communication over AWGN channels. \footnote{Part of this work has been presented at Allerton 2019 \cite{Yu2}.}
\end{abstract}
\section{Introduction}
The broadcast nature of wireless communications makes the security of communication through it an acute matter. Covert communication, different from typical secret communication, has earned much attention in recent years. In this  circumstance, the adversary should have a low probability of detection (LPD) of the transmitted message. Such scenarios arise in underwater acoustic communication \cite{Roee Diamant} and dynamic spectrum access in wireless channels, where secondary users attempt to communicate without being detected by primary users or users wish to avoid the attention of regulatory entities \cite{Matthieu R}. The information theory for the low probability detection communication was first characterized on AWGN channels in \cite{Boulat A} and DMCs in \cite{Matthieu R}\cite{Ligong Wang}, and later in \cite {Pak Hou Che} and \cite{Abdelaziz} on BSC and MIMO AWGN channels, respectively. It has been shown that LPD communication follows the following square root law.
\begin{Square Root Law}
In covert communication,  for any $\varepsilon > 0$, the transmitter is able to transmit $O(\sqrt{n})$ information bits to the legitimate receiver by $n$ channel uses while lower bounding the adversary's sum of probability of detection errors $\alpha + \beta \geq 1- \varepsilon$ if she knows a lower bound of the adversary's noise level ($\alpha$ and $\beta$ are error probabilities of type I and type II in the adversary's hypothesis test). The number of information bits will be $o(\sqrt{n})$ if she doesn't know the lower bound.
\end{Square Root Law}
A number of works focused on improving the communication efficiency by various means, such as using channel uncertainty in \cite{Seonwoo}\cite{Biao He}\cite{Khurram Shahzad}, using jammers in \cite{Sobers}\cite{Tamara V.Sobers} and other methods in \cite{B.A.Bash2}\cite{Soltani}. These methods are discussed in the asymptotic regime.
However, in practical communication, we are more concerned about the behaviors in finite blocklength regime. For example, given a finite block length $n$, how many information bits can be transmitted with a given covert criterion and maximal probability of error $\epsilon$, under which the adversary is not able to determine whether or not the transmitter is communicating effectively. When the channels are discrete memoryless, this question has been answered by \cite{M.Tahmasbi}\cite{M.Tahmasbi2}, where the exact second-order asymptotics of the maximal number of reliable and covert bits are characterized when the discrimination metrics are relative entropy, total variation distance (TVD) and missed detection probability with fixed probability of false alarm, respectively. For AWGN channels and slow fading channels, the maximal transmit power and the maximal transmit bits in finite block length are partly characterized in \cite{Yan1} and \cite{Haoyue Tang}, respectively by applying the results in finite blocklength regime. However, it is inappropriate to directly apply the bounds in \cite{Polyanskiy} for Gaussian signaling without proper discretion since the power constraint is not satisfied with probability $1$ for Gaussian signalling in finite blocklength regime.

In this work, we pose a problem of covert communication in non-asymptotic scenario: if Gaussian codewords are utilized under a covert constraint in the form of TVD, how much throughput shall we expect? Is it possible to give bounds in a concise form making them relatively easy to evaluate? In literature, Kullback-Leibler divergence, rather than TVD, has often be adopted as a metric of covertness \cite{Boulat A}\cite{Ligong Wang} to quantify covertness of Alice's transmission schemes. We are interested in covert constraint in the form of TVD because it has range $[0,1]$, hence is a normalized metric of discrimination for two probability measures. Moreover, it does not increase with the blocklength (KL divergence will increase linearly with blocklength if i.i.d random codewords are adopted.) and is directly related to the effect of hypothesis testing at the adversary. These advantages prompt us to choose an upper bound of TVD as a covert constraint in the finite blocklength regime. There are two reasons for us to be interested in Gaussian codewords. First, Gaussian distribution has its advantage of both maximizing the mutual information between the input and output ends of the legitimate receiver over AWGN channels in the asymptotic regime and minimizing KL divergence between the output and the background noise at the adversary \cite{Ligong Wang}. It has found applications in secure chaotic spread spectrum communication systems \cite{Alan1}\cite{Alan2}. Second, the total variation distance at the adversary is relatively easy to analyze when the codewords are Gaussian generated (or nearly Gaussian generated), which has its advantage over a determined codebook where the discrimination at the adversary is difficult to handle analytically in general. In addition, random coding approach, which includes deterministic coding
as a special case, can offer us means to attain even greater achievability bounds on the number of decodable codewords
since the distribution of the codewords is at our disposal. In \cite{Polyanskiy}, the achievability bound over AWGN channel is obtained from a deterministic codebook on the surface of n-dimensional sphere with radius $\sqrt{nP}$ where $P$ is fixed. The sphere symmetry simplifies the calculation by considering a particular codeword with equal coordinates. In contrast with that, a randomly generated codeword of length $n$ following a Gaussian distribution will have probability of zero to attain a pre-specified $L_2$ norm. Moreover, TVD constraint will lead to maximal power constraint on each codeword. As a result, the sample space of the vectors will be only a subset of n-dimensional space. The dilemma has prompted us to go through carefully and cautiously  the techniques  developed in \cite{Polyanskiy} in order to establish achievability bounds for AWGN channels with Gaussian input and maximum power constraints. In fact, we are mainly interested in providing achievability and converse bounds of covert communication and their normal approximations which are convenient to evaluate. It is not our main concern to derive bounds which outperform the existing ones.  Nevertheless, it is essential for us to point out the necessary revisions and changes in accordance with finite blocklength and covert constraint as a specialized scenario different from \cite{Polyanskiy}.  In particular, to develop new bounds suitable for covert communication over AWGN channels, we need to carefully integrate such techniques including random generation of codewords, and binary hypothesis testing at decoder side.  More specifically, the major and novel contributions of our work are listed below.
\begin{itemize}
\item New achievability bounds (with both maximal probability of error and average probability of error) are obtained for cases with random coding and input constraints.
\item New converse and achievability bounds are found on the channel coding rate over AWGN channels when the codewords are generated from Gaussian distribution and selected from a set with maximal power constraint.
\item Normal approximations for both bounds over covert channel are presented. Morover, we provide detailed discussions on the difference between our achievability bounds and existing ones.
\item The optimal distributions for both bounds are investigated and we show that they are solutions to the two minimax
     and maxmin optimization problems posed with respect to TVD metrics. These facts shed light on the optimal
     coding schemes to obtain larger coding rate in the finite blocklength regime.
\item The TVD at the adversary under the coding scheme is analyzed in details. Some divergence inequalities are applied to ensure the TVD at the adversary is controlled at a given threshold with proper chosen parameters.
\item The attainable throughputs of covert communications under such TVD constraints, as well as the error rates in the finite block length regime are evaluated numerically in a variety of situations.
\end{itemize}

The main differences of this paper from \cite{Yu2} includes the comparison between our bounds and the previous ones, the identified relationships between the optimal channel input distributions and the posed minimax and maxmin problems for attaining achievability and converse bounds, respectively, and the analysis on TVD and the throughput in covert communication.

The rest of this paper is arranged as follows. In Section II, we describe the model for covert communication over AWGN channels. In Section III, general results on achievability bound are provided. The main results over AWGN channel are presented in Section IV and Section V. The applications of the bounds under covert constraints are presented in Section VI. Finally, the conclusion is drawn in Section VII.

\section{The Channel Model}\label{model}

In this section, the channel model of covert communication over AWGN channels is presented.
An $(n,2^{nR})$ code for the Gaussian covert communication channel consists of a message set $W \in \mathcal{W}=\{1,...,2^{nR}\}$, an encoder at the transmitter Alice $f_n : \mathcal{W}\rightarrow \mathbb{R}^n, w\mapsto x^n$, and a decoder at the legitimate user Bob $g_n : \mathbb{R}^n \rightarrow \mathcal{W}, y^n \rightarrow \hat{w}$. Meanwhile, a detector is at an adversary Willie $h_n : \mathbb{R}^n \rightarrow \{0,1\}, z^n \rightarrow 0/1$. The error probability of the code is defined as $P^n_e =Pr[g_n (f_n(W))\neq W]$.
\begin{figure}
\centering
\includegraphics[width=2.5in]{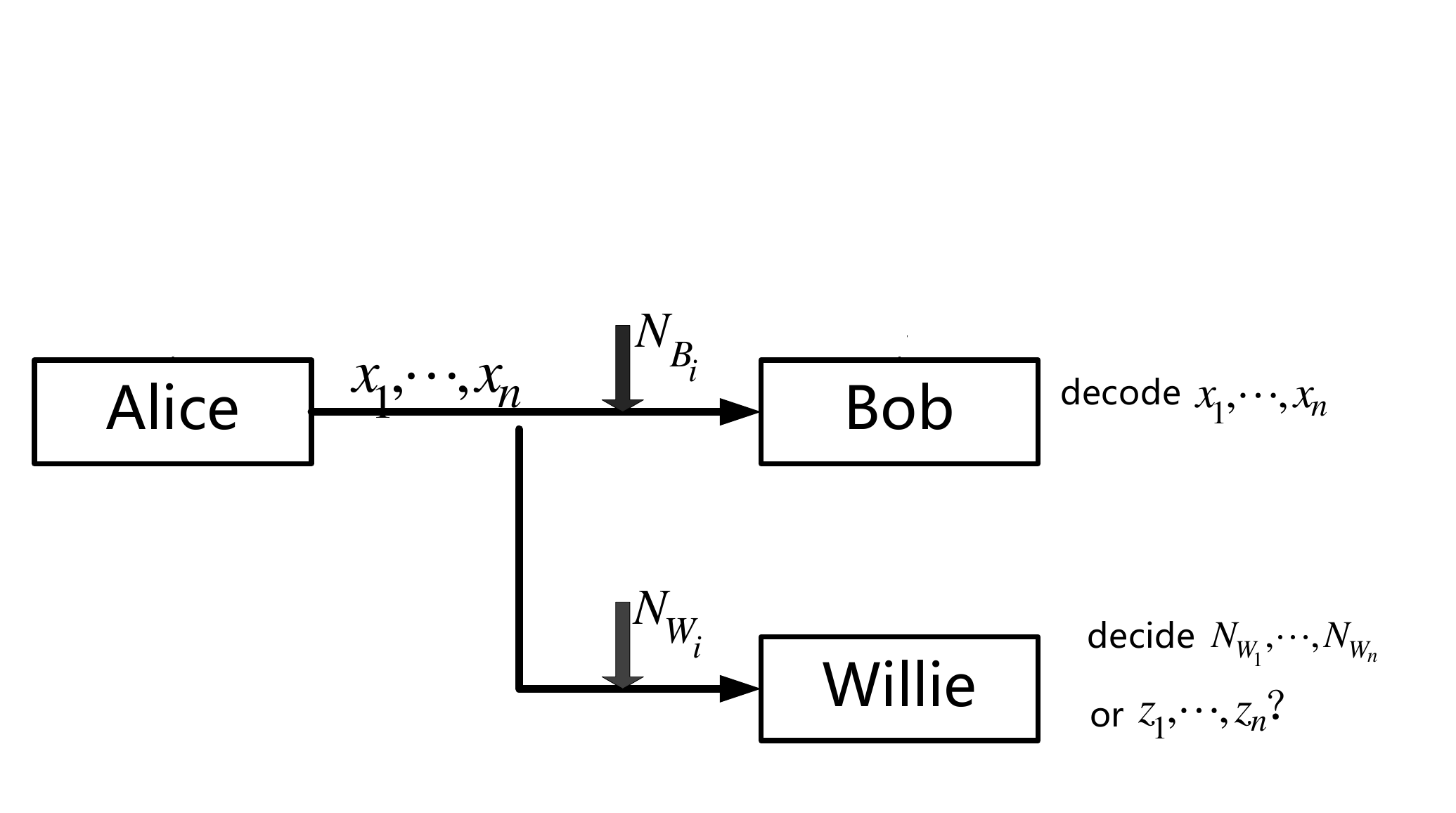}
\caption{The channel model of Gaussian LPD communication in Section \ref{model}}\label{Fig1}
\end{figure}
As shown in Fig.\ref{Fig1}, the channel model is defined by
\begin{eqnarray}
y_i=x_i+N_{B_i},i = 1,...,n, \label{channel11}\\
z_i=x_i+N_{W_i},i = 1,...,n,  \label{channel2}
\end{eqnarray}
where $x^n=\{x_i\}_{i=1}^n$, $y^n=\{y_i\}_{i=1}^n$ and $z^n=\{z_i\}_{i=1}^n$ denote Alice's input codeword, the legitimate user Bob's observation and the adversary Willie's observation, respectively. $N_{B_i},i=1,...n$, is  independent identically distributed (i.i.d) according to $\mathcal{N}(0,\sigma_b^2)$. The quantity $N_{W_i},i=1,...n$, is independent of $N_{B_i}$ and is i.i.d according to $\mathcal{N}(0,\sigma^2_w)$. For convenience, it is assumed that $\sigma^2_w = \sigma^2_b = 1$. Each codeword is randomly selected from a subset of candidate codewords. Each coordinate of these candidates are i.i.d generated from $\mathcal{N}(0,P(n))$ where $P(n)$ is a decreasing function of $n$. The detail of selection will be discussed later. The adversary is aware that the codebook is generated from Gaussian distribution $\mathcal{N}(0, P(n))$ with blocklength $n$ but he doesn't know the specific codebook.
The adversary Willie tries to determine whether Alice is communicating ($h_n=1$) or not ($h_n=0$) by statistical hypothesis test. Alice, who is active about her choice, is obligated to seek for a code such that $\lim_{n\rightarrow \infty}P^n_e\rightarrow 0$ and $\lim_{n\rightarrow \infty}P(h_n=0)\rightarrow \frac{1}{2}$. There is usually a secret key to assist the communication between Alice and Bob (such as the identification code for the users in spread spectrum communication), which is not the focus of this work. The interested reader may refer to \cite{Matthieu R} and \cite{Boulat A} for more details.

The hypothesis test of Willie in covert communication is performed on his received signal $z^n$ which is a sample of random vector $Z^n$. The null hypothesis $H_0$ corresponds to the situation that Alice doesn't transmit and consequently $Z^n$ has output probability distribution $\mathbb{P}_0$.
Otherwise, the received vector $Z^n$ has output probability distribution $\mathbb{P}_1$ which depends on the input distribution. The rejection of $H_0$ when it is true will lead to a false alarm with probability $\alpha$.
The acceptance of $H_0$ when it is false is considered to be a miss detection  with probability $\beta$. The aim of Alice is to decrease the success probability of Willie's test by increasing $\alpha + \beta$, and meanwhile obtain reliable communication with Bob. The effect of the optimal test is usually measured by the total variation distance (TVD) $V_T(\mathbb{P}_1, \mathbb{P}_0)$
 which is $1-(\alpha+ \beta)$ (Theorem 13.1.1 in \cite{E.Lehmann}).
 The total variation distance between two probability measures $P$ and $Q$ on a sigma-algebra $\mathcal {F}$ of subsets of the sample space $\Omega$ is defined as
 \begin{equation}\label{TVDD}
 \begin{split}
  V_T(P,Q)= \underset{A \in \Omega}{\sup} \left|P(A)-Q(A)\right|.
 \end{split}
 \end{equation}
When $V_T(\mathbb{P}_1, \mathbb{P}_0)$ is close to $0$, it is generally believed that any detector at Willie can not discriminate the induced output distribution and the distribution of noise effectively, hence can not distinguish whether or not Alice is communicating with Bob.
  Our interest is to find both achievability and converse bounds in the finite block regime with any given $n$ under a constraint of an imposed upper-bound on TVD between two Gaussian distributions (More accurately, the distribution of noise is Gaussian, and the output distribution induced by the code is not Gaussian in the strict sense). As a result of such bound on TVD, the transmission power $P(n)$ of Alice is a decreasing function of $n$. In asymptotic situation, it is shown in \cite{Ligong Wang} (Theorem 5) that Gaussian codebook is optimal over AWGN channel for covert constraint $D(Q^n\|Q^{\times n}_0) \leq \delta$ where $Q^n$ is the induced output distribution and $Q^{\times n}_0$ is the distribution of background noise. When the variance of noise distribution is $1$, each codeword is independent generated with i.i.d coordinates from $\mathcal{N}(0, P(n))$:
 \begin{equation}
 P(n) = 2 \sqrt{\frac{\delta}{n}}.
 \end{equation}
 In this work, to satisfy the covert constraint in the form of TVD: $V_T(\mathbb{P}_1, \mathbb{P}_0)\leq \delta$, proper $P(n)$ as a decreasing function of finite $n$ should also be determined. Before diving deep into the details of covert constraint, we establish the main framework of new coding scheme and corresponding one-shot bounds in next two sections. The choice of $P(n)$ and some other parameters in covert channel will be investigated in Section VI.

\section{General Results on Achievability Bound under Maximal Power Constraint}\label{A1}
\subsection{Preliminary}
In this section, we introduce primary definitions used through the rest of the paper.  Let two sets $\mathsf{A}$ and $\mathsf{B}$ be input and output sets of a communication system with conditional probability measure $P_{Y|X} : \mathsf{A} \mapsto \mathsf{B}$. A codebook is a set of codewords $(c_1, \cdots, c_M) \in \mathsf{A}^M$. An encoder is a function from $[M] = \{1,\cdots, M\}$ to $\mathsf{A}^M$: $ W \mapsto c_W $ and the decoder is defined as $P_{\hat{W}|Y}: \mathsf{B} \mapsto \{0, 1, \cdots,M\}$ (here `$0$' indicates ``error") where $\hat{
W}$ is a random variable representing the index of corresponding message or an error.

There are two kinds of metric for the error probability to judge the quality of a code, i.e., average error probability and maximal error probability, which are defined as follows:
\begin{enumerate}
\item [i]
\begin{equation}
P_e \triangleq \mathbb{P}\left[W \neq \hat{W}\right];
\end{equation}
\item [ii]
\begin{equation}
P_{e,max} \triangleq \max_{m \in [M]}\mathbb{P}\left[\hat{W} \neq m | W = m\right].
\end{equation}
\end{enumerate}
When a codebook and its decoder satisfies $P_e \leq \epsilon$ (or $P_{e,max} \leq \epsilon$), they are called an $(M, \epsilon)$ code with average error probability $\epsilon$ (or maximal error probability).
For a joint distribution $P_{XY}$ on $\mathsf{A}\times \mathsf{B}$, the information density is
\begin{equation}
i(x;y) = \log \frac{dP_{Y|X=x}}{dP_Y}(y).
\end{equation}
When $P_{Y|X=x}$ is not absolutely continuous with respect to $P_Y$, the information density is defined to be $+\infty$ (or $- \infty$) if $y$ is in the singular set (or $\{y: \frac{dP_{Y|X=x}}{dP_Y}(y) = 0\}$).
\subsection{Achievability Bounds in General Settings}
As Gaussian random coding will be involved, we should rely on the finite blocklength bound of random coding scheme. However, the existing results in the literature, such as the bounds provided in (108) and (127) in \cite{Polyanskiy} cannot be directly employed in our scenario because they are based on deterministic coding, not random coding considered here. In order to get achievability results for our scenario, we will construct a coding scheme based on Part C of Section III in \cite{Polyanskiy}. The code is randomly and sequentially constructed step by step by random coding argument. The decoding procedure of the code is determined as sequential dependence testing. Our results are based on combined application of the following elements.
\begin{itemize}
\item The codewords are randomly chosen from a set $\mathsf{F}$ which is a subset induced by some constraint on the whole space $\mathsf{A}$.
\item For each codeword $x$, there is an associated threshold $\gamma(x)$ in the dependence testing.
\end{itemize}
\begin{Lemma}\label{Le1}
 For any distribution $P_X$ on $\mathsf{A}$, and any measurable function $\gamma: \mathsf{A} \mapsto [0, \infty]$, there exists a code with $M$ codewords in the set $\mathsf{F}$ whose maximal error probability satisfies
\begin{equation}
\begin{split}
\epsilon P_X[\mathsf{F}] &\leq E_X[\mathbb{P}(i(x;Y) \leq \log\gamma(x))\cdot 1_{\{x\in \mathsf{F}\}}] \\
+ &(M-1)P_X[\mathsf{F}] \cdot \underset{x\in \mathsf{F}}{\sup}P_Y[i(x;Y) > \log \gamma(x)].
\end{split}
\end{equation}
\end{Lemma}
\begin{Proof}

The operation of the decoder is the following sequential decoding process. It computes $i(c_j;y)$ for the received channel output $y$ and selects the first codeword $c_j$ who satisfies $i(c_j; y) > \log \gamma(c_j)$.
For the first codeword, the conditional probability of error under the decoding rule is
\begin{equation}
\epsilon_1(x) = \mathbb{P}[i(x;Y)\leq \log \gamma(x)| X = x]
\end{equation}
once the codeword $x$ is chosen.
Since $x$ is chosen from $\mathsf{F}$, we have
\begin{equation}
\begin{split}
\epsilon_1(c_1) \leq& \mathbb{E}[\epsilon_1(x)|\mathsf{F}]\\
= & \underset{x\in \mathsf{F}}{\sum}\mathbb{P}(i(x;Y) \leq \log \gamma(x) |X=x)P(X=x|\mathsf{F})\\
= & \underset{x\in \mathsf{F}}{\sum}\frac{P(X=x, x\in \mathsf{F})}{P_X[\mathsf{F}]}\mathbb{P}[i(x;Y)\leq \log \gamma(x)]\\
= &\frac{\underset{x\in \mathsf{F}}{\sum}P(X = x)\mathbb{P}[i(x;Y)\leq \log \gamma(x)]}{P_X[\mathsf{F}]}.
\end{split}
\end{equation}
If we assume that $j-1$ codewords $\{c_l\}_{l=1}^{j-1}$ have been chosen, denote
\begin{equation}
D_{j-1} = \bigcup_{l=1}^{j-1}\{y: i(c_l;y) > \log \gamma(c_l)\} \subseteq \mathnormal{B}.
\end{equation}
The conditional probability of error that $x \in \mathsf{F}$ is chosen to be the $j$th codeword is
\begin{equation}\label{error}
\epsilon_j(c_1, \cdots, c_{j-1}) = 1 - \mathbb{P}[{i(x;Y)> \log \gamma(x)}\backslash D_{j-1}|X=x].
\end{equation}
The expectation of error is actually the conditional expectation as follows,
\begin{equation}\label{imp}
\begin{split}
&\mathbb{E}[\epsilon_j(c_1,\cdots,c_{j-1},X)|\mathsf{F}] \\
= &\underset{x\in \mathsf{F}}{\sum}\mathbb{P}[{i(x;Y)\leq \log \gamma(x)}\cup D_{j-1}|x]P(X = x| \mathsf{F})\\
\leq & \underset{x\in \mathsf{F}}{\sum}\mathbb{P}[{i(x;Y)\leq \log \gamma(x)}|x]P(X = x| \mathsf{F}) + P_Y(D_{j-1})\\
\leq &\frac{\underset{x\in \mathsf{F}}{\sum}P(X = x)\mathbb{P}[i(x;Y)\leq \log \gamma(x)]}{P_X[\mathsf{F}]} \\
+ &(j-1)\underset{x\in \mathsf{F}}{\sup}P_Y[i(x;Y) > \gamma(x)].
\end{split}
\end{equation}
The first equality is from the fact that $ \{{i(x;Y)> \log \gamma(x)}\cap D_{j-1}^C\}^C = \{i(x;Y)\leq \log \gamma(x)\}\cup D_{j-1}$, and the expectation of (\ref{error}) is less than
\begin{equation}\label{error2}
\begin{split}
\underset{x\in \mathsf{F}}{\sum}P_{Y|X=x}[{i(x;Y)\leq \log \gamma(x)}|x]P(X = x| \mathsf{F}) \\
+ \underset{x\in \mathsf{F}}{\sum}P_{Y|X=x}(D_{j-1})\cdot P(X=x|\mathsf{F})
\end{split}
\end{equation}
by union bound.
The second term of (\ref{error2}) can be rewritten as
\begin{equation}\label{Fu}
\begin{split}
&\underset{x\in \mathsf{F}}{\sum}P_{Y|X=x}(D_{j-1})\cdot P(X=x|\mathsf{F})\\
= &\underset{x\in \mathsf{F}}{\sum}\underset{y\in D_{j-1}}{\sum}P_{Y|X=x}(y)P(X=x|\mathsf{F})\\
= &P_Y(D_{j-1}).
\end{split}
\end{equation}
The last equality of (\ref{Fu}) is from the fact that the induced output distribution $P_Y$ has probability mass function
\begin{equation}
P_Y(y) = \underset{x\in \mathsf{F}}{\sum}P_{Y|X=x}(y)\cdot P(X=x|\mathsf{F})
\end{equation}
and Fubini's Theorem.

Note that we usually use $\mathbb{P}$ for the particular induced distribution $P_{Y|X=x}$. Thus, there exists a codeword $c_j \in \mathsf{F}$ such that $\epsilon_j(c_1, \cdots, c_{j-1},c_j)$ satisfies
\begin{equation}
\begin{split}
\epsilon_j P_X[\mathsf{F}] \leq &\underset{x\in \mathsf{F}}{\sum}P(X = x)\mathbb{P}[i(x;Y)\\
\leq &\log \gamma(x)] + (j-1)P_X[\mathsf{F}]\cdot\underset{x\in \mathsf{F}}{\sup}P_Y[i(x;Y) > \gamma(x)].
\end{split}
\end{equation}
In particular, the maximal error probability should satisfy
\begin{equation}
\begin{split}
\epsilon P_X[\mathsf{F}] &\leq \underset{x\in \mathsf{F}}{\sum}P(X = x)\mathbb{P}[i(x;Y)\leq \log \gamma(x)]\\
 + &(M-1)\underset{x\in \mathsf{F}}{\sup}P_Y[i(x;Y) > \gamma(x)]\\
= &E_X[\mathbb{P}(i(x;Y) \leq \log\gamma(x))\cdot 1_{\{x\in \mathsf{F}\}}] \\
+& (M-1) P_X[\mathsf{F}]\cdot\underset{x\in \mathsf{F}}{\sup}{P_Y[i(x;Y) > \log \gamma(x)]}.
\end{split}
\end{equation}
\end{Proof}
\begin{Remarks}
\hspace{0.5cm}
\begin{itemize}
\item This lemma is different from Theorem 23 in \cite{Polyanskiy} from two aspects. The first term of the right side of inequality is summed over the subset $\mathsf{F}$ but not the whole set $\mathsf{A}$. In addition, $P_Y$ is the induced unconditional distribution of the codewords whose support is $\mathsf{F}$, which is different from the induced distribution $P_Y$ of Theorem 23.
\item There is a $P_X[\mathsf{F}]$ in the second term of the right side of the inequality, which is missing in \cite{Yu2}.\footnote{In step (12) of Lemma 1 in \cite{Yu2}, $P_X[\mathsf{F}]$ should be multiplied on the second term of the right side. The update is also applicable for Lemma 2, Theorem 1 and Theorem 2.} Though it has little influence on the subsequent analysis and the main results, we add it for mathematical rigor.
\item The nature of the construction is as follows. We have a distribution $P_X$ which is easy for us to generate codewords, but we just want the codewords in a subset $\mathsf{F}$, so we truncate $P_X$ and re-normalize it to get a distribution $\bar{P}_X$ concentrated in $\mathsf{F}$. Then we use the construction of Theorem 21 in \cite{Polyanskiy} on base of $\bar{P}_X$.
\end{itemize}
\end{Remarks}

\begin{Lemma}\label{Le2}
For any distribution $P_X$ on $\mathsf{A}$, and any measurable function $\gamma: \mathsf{A} \mapsto [0, \infty]$, there exists a code with $M$ codewords in the set $\mathsf{F}$ whose average error probability satisfies
\begin{equation}
\begin{split}
\epsilon P_X[\mathsf{F}] &\leq E_X[\mathbb{P}(i(x;Y) \leq \log\gamma(x))\cdot 1_{\{x\in F\}}] \\
+ &\frac{M-1}{2} P_X[\mathsf{F}]\cdot \underset{x\in \mathsf{F}}{\sup}{P_Y\left[i(x;Y) > \log \gamma(x)\right]}.
\end{split}
\end{equation}
\end{Lemma}
\begin{Proof}
As we have shown in Lemma \ref{Le1} that there exists a codebook $\{c_j\}_{j=1}^M$, the conditional error probability given the $j$th codeword satisfies
\begin{equation}
\begin{split}
\epsilon_j \leq \frac{\underset{x\in \mathsf{F}}{\sum}P(X = x)\mathbb{P}[i(x;Y)\leq \log \gamma(x)]}{P_X[\mathsf{F}]}\\
+ (j-1) \underset{x\in \mathsf{F}}{\sup}P_Y[i(x;Y) > \gamma(x)].
\end{split}
\end{equation}
As the codewords are equiprobable and the average error probability satisfies
\begin{equation}
\begin{split}
\epsilon \leq &\frac{\underset{x\in \mathsf{F}}{\sum}P(X = x)\mathbb{P}[i(x;Y)\leq \log \gamma(x)]}{P_X[\mathsf{F}]}\\
 + &\frac{M-1}{2}\underset{x\in \mathsf{F}}{\sup}P_Y[i(x;Y) > \gamma(x)].
 \end{split}
\end{equation}
Consequently, we have
\begin{equation}
\begin{split}
\epsilon P_X[\mathsf{F}] \leq & \underset{x\in \mathsf{F}}{\sum}P(X = x)\mathbb{P}[i(x;Y)\leq \log \gamma(x)] \\
+ &\frac{M-1}{2} P_X[\mathsf{F}]\cdot \underset{x\in \mathsf{F}}{\sup}P_Y[i(x;Y) > \gamma(x)]\\
\end{split}
\end{equation}
\end{Proof}

\begin{Remark}
The above lemma is in general weaker than the next one. However, it is more convenient to evaluate when the computation of the expectation of $P_Y[i(x;Y) > \log\gamma(x)]$ over $x \in \mathsf{F}$ is difficult.
\end{Remark}
\begin{Lemma}\label{Le3}
 For any distribution $P_X$ on $\mathsf{A}$, and any measurable function $\gamma: \mathsf{A} \mapsto [0, \infty]$, there exists a code with $M$ codewords in the set $\mathsf{F}$ whose average error probability satisfies
\begin{equation}\label{aver2}
\begin{split}
\epsilon P_X[\mathsf{F}] \leq & E_X[\mathbb{P}(i(x;Y) \leq \log\gamma(x))\cdot 1_{\{x\in \mathsf{F}\}}] \\
+ &\frac{M-1}{2} E_X[P_Y(i(x;Y) > \log \gamma(x))\cdot 1_{\{x\in \mathsf{F}\}}].
\end{split}
\end{equation}
\end{Lemma}
\begin{Proof}
In the step (\ref{imp}) in the proof of Lemma \ref{Le1}, we rewrite it as
\begin{equation}
\begin{split}
&\mathbb{E}[\epsilon_j(c_1,\cdots,c_{j-1},X)|\mathsf{F}] \\
= &\underset{x\in \mathsf{F}}{\sum}\mathbb{P}[{i(x;Y)\leq \log \gamma(x)}\cup D_{j-1}|x]P(X = x| \mathsf{F})\\
\leq & \underset{x\in \mathsf{F}}{\sum}\mathbb{P}[{i(x;Y)\leq \log \gamma(x)}|x]P(X = x| \mathsf{F}) + P_Y(D_{j-1})\\
 \leq & \underset{x\in \mathsf{F}}{\sum}\mathbb{P}[{i(x;Y)\leq \log \gamma(x)}|x]P(X = x| \mathsf{F})\\
  + &\sum_{l=1}^{j-1} P_Y[i(c_l; Y) > \gamma(c_l)].
\end{split}
\end{equation}
The last inequality is from  union bound.
Thus, we have there exists a codeword $c_j \in \mathsf{F}$ which satisfies
\begin{equation}\label{Exist}
\begin{split}
\epsilon_j \leq \underset{x\in \mathsf{F}}{\sum}\mathbb{P}[{i(x;Y)\leq \log \gamma(x)}|x]P(X = x)\\
 + \sum_{l=1}^{j-1} P_Y[i(c_l; Y) > \gamma(c_l)]
\end{split}
\end{equation}
Note that the order of the index does not affect both two terms of (\ref{Exist}). For the second term, consider all possible orders of the indexes, we have there exists an order such that the union bound satisfies
 \begin{equation}
 \begin{split}
 &\sum_{l=1}^{j-1} P_Y[i(c_l; Y) > \gamma(c_l)] \\
 \leq & (j-1)\cdot E_X[P_Y(i(x;Y) > \gamma(x))|\mathsf{F}]\\
  \leq & (j-1)\cdot \underset{x\in \mathsf{F}}{\sum}P_Y[i(x;Y)> \log \gamma(x)]P(X=x|\mathsf{F}).
 \end{split}
 \end{equation}
Thus, we have
 \begin{equation}\label{Exist1}
 \begin{split}
\epsilon_j \leq &\, \ \,\underset{x\in \mathsf{F}}{\sum}\mathbb{P}[{i(x;Y)\leq \log \gamma(x)}|x]P(X = x)\\
 &+ (j-1)\cdot \underset{x\in \mathsf{F}}{\sum}P_Y[i(x;Y)> \log \gamma(x)]P(X=x|\mathsf{F}).
 \end{split}
\end{equation}

As the codewords are equiprobable, the average probability of error should satisfy
\begin{equation}
\begin{split}
\epsilon &\leq \frac{\underset{x\in \mathsf{F}}{\sum}P(X = x)\mathbb{P}[i(x;Y)\leq \log \gamma(x)]}{P_X[\mathsf{F}]}\\ + &\frac{1}{M}\sum_{j=1}^M(j-1)\underset{x\in \mathsf{F}}{\sum}P_Y[i(x;Y)> \log \gamma(x)]P(X=x|\mathsf{F})\\
& \leq \frac{\underset{x\in \mathsf{F}}{\sum}P(X = x)\mathbb{P}[i(x;Y)\leq \log \gamma(x)]}{P_X[\mathsf{F}]} \\
+ &\frac{M-1}{2}\frac{E_X[P_Y(i(x;Y) > \log \gamma(x))\cdot 1_{\{x\in F\}}]}{P_X[\mathsf{F}]}.
\end{split}
\end{equation}
Thus, we have proved (\ref{aver2}).
\end{Proof}

Note that in the above results, the real distribution of the choosing codewords is a truncated distribution of $P_X$, and $P_Y$ is the induced output distribution of the truncated distribution.

\subsection{Further Results on Achievability Bounds}
Now consider the binary hypothesis test between $P_{Y|X = x}$ and the induced output distribution $P_Y$ on $\mathsf{W}$. Let us introduce the detection probability
\begin{equation}
\begin{split}
P_{Y|X=x}[i(x;Y) > \log \gamma(x)]
 \geq 1 - \epsilon + \tau(x)
\end{split}
\end{equation}
 with $ 0 < \tau(x) < \epsilon$
of Neyman-Pearson hypothesis tests with decision threshold $\log \gamma(x)$ when the sending codeword is $x$. The details of Neyman-Pearson hypothesis testing can be found in Appendix B of \cite{Polyanskiy}.
Note that for a particular codeword $x$, we do have a separate threshold $\gamma(x)$, and the resulting false alarm probability for this particular $x$ is
$P_Y  \left[ i(x,Y) > \log \gamma(x) \right] = \beta_{1-\epsilon + \tau(x)}$.
It is obvious that a lower bound on the decision threshold $\gamma(x)$ is equivalent as a lower bound on the detection probability $1-\epsilon + \tau(x)$.

The following theorem is a combination of random coding, and selecting $\gamma(x)$ for each $x$ in a subset $\mathsf{F}$ such that for each  $x$ in this set, we have the detection probability in favor of $P_{Y|X=x}$ over $P_Y$  lower bounded by $1-\epsilon+\tau(x)$ with an additional constraint: $\tau(x)< \epsilon$ for all $x \in \mathsf{F}$.

\begin{Theorem}\label{AcTheorem1}
For any input distribution $P_X$ on $\mathsf{A}$ and measurable function $\tau: \mathsf{A}\rightarrow [0, \infty]$, there exists a code with $M$ codewords in the set $\mathsf{F} \subseteq \mathsf{A}$ such that the maximal error probability $\epsilon$ satisfies
\begin{equation}\label{aac1}
M \geq \frac{E_X\left[\tau(x)\cdot 1_{\{x\in \mathsf{F}\}} \right]}{P_X[\mathsf{F}] \cdot \underset{x\in \mathsf{F}}{\sup}\beta_{1-\epsilon+ \tau(x)}(x, P_Y)}
\end{equation}
where $\beta_{1-\epsilon+ \tau(x)}(x,P_Y)$ is the minimum probability of error under hypothesis $P_Y$ if the probability of error when $x$ is sent is not larger than $1-\epsilon+\tau(x)$.
\end{Theorem}
\begin{Proof}
The theorem is an application of Lemma \ref{Le1} with dependent testing is substituted by Neyman-Pearson tests.
We have
\begin{equation}
M \geq \frac{\epsilon P_X[\mathsf{F}] - E_X[\mathbb{P}(i(x;Y) \leq \log\gamma(x))\cdot 1_{\{x\in \mathsf{F}\}}]}{P_X[\mathsf{F}] \cdot \underset{x\in \mathsf{F}}{\sup}{P_Y[i(x;Y) > \log \gamma(x)]}}
\end{equation}
from Lemma \ref{Le1}.
As $P_Y[i(x;Y) > \log \gamma(x)] = \beta_{1 - \epsilon + \tau(x)}$ from above analysis, then
$P_{Y|X=x}[i(x;Y) \leq \log \gamma(x)] \leq \epsilon - \tau(x)$. The conclusion is obvious.
\end{Proof}
\begin{Corollary}\label{AcCorollary1}
If the codewords are chosen from $\bar{\mathsf{F}}$, which is the subset of $\mathsf{F}$ that satisfies $\tau(x) \geq \tau_0$, the bound with maximal error probability $\epsilon$ can be rewritten as
\begin{equation}\label{bound20}
M \geq \underset{0 < \tau_0 < \epsilon}{\sup}\frac{\tau_0}{\underset{x\in \bar{\mathsf{F}}}{\sup}\beta_{1-\epsilon+ \tau(x)}(x, P_Y)}.
\end{equation}
If we let $\tau(x) = \tau_0$, then the above bound is
\begin{equation}\label{bound2}
M \geq \underset{0 < \tau_0 < \epsilon}{\sup}\frac{\tau_0}{\underset{x\in \bar{\mathsf{F}}}{\sup}\beta_{1-\epsilon+ \tau_0}(x, P_Y)}.
\end{equation}
\end{Corollary}
Note that these bounds look like (127) in \cite{Polyanskiy}, the underlying code construction is random, though.

The following two achievability bounds of average error probability are direct applications of Lemma \ref{Le2} and Lemma \ref{Le3}.
\begin{Theorem}
For any distribution $P_X$ on set $\mathsf{A}$ there exists a code with $M$ codewords in $\mathsf{F}$ and average probability of error satisfying
\begin{equation}\label{acc2}
M \geq \frac{2\cdot E_X\left[\tau(x)\cdot 1_{\{x\in \mathsf{F}\}} \right]}{P_X[\mathsf{F}]\cdot\underset{x\in \mathsf{F}}{ \sup}\beta_{1-\epsilon+\tau(x)}(x, P_Y)}.
\end{equation}
\end{Theorem}
\begin{Theorem}\label{AVER}
For any distribution $P_X$ on set $\mathsf{A}$ there exists a code with $M$ codewords in $\mathsf{F}$ and average probability of error satisfying
\begin{equation}\label{aver3}
\frac{M-1}{2} \geq \frac{E_X\left[\tau(x)\cdot 1_{\{x\in \mathsf{F}\}} \right]}{E_X\left[\beta_{1-\epsilon+ \tau(x)}(x, P_Y) \cdot 1_{\{x\in \mathsf{F}\}}\right]}.
\end{equation}
\end{Theorem}

\begin{Remarks}
\hspace{0.5cm}
\begin{itemize}
\item It is obvious that (\ref{aver3}) is better than (\ref{acc2}) due to the advantage from average operator. Thus, random coding actually shall provide us more leverage in attaining a better achievability bound. Moreover, we can even
consider deterministic coding as a particular way of random coding in that the distribution of the codebook is
concentrated over a specific one.
\item In above results, $\tau(x)$ is a function of the randomly selected codewords, which depends on the specific $x$. It should be within $(0, \epsilon)$ and depends on the encoding scheme and $x$. Moreover, any function which satisfies $\tau(x) \in (0, \epsilon)$ should be feasible.  Tighter bounds can be obtained by searching the optimal function $\tau(x)$ to maximize the right-hand side of all these bounds. For example, the bound in (\ref{aac1}) can be further optimized as
    \begin{equation}\label{aac2}
M \geq \underset{\tau(x)}{\sup}\frac{E_X\left[\tau(x)\cdot 1_{\{x\in \mathsf{F}\}} \right]}{P_X[\mathsf{F}]\cdot\underset{x\in \mathsf{F}}{\sup}\beta_{1-\epsilon+ \tau(x)}(x, P_Y)}.
\end{equation}
The same optimization can be applied on (\ref{aver3}).
\item
In Theorem \ref{AVER}, if we use the subset $\bar{\mathsf{F}}$ with constraint that $\tau(x) > \tau_0$  instead and ${\underset{x\in \bar{\mathsf{F}}}{\sup}\beta_{1-\epsilon+ \tau(x)}(x, P_Y)}$ in the denominator instead of the average over $\bar{\mathsf{F}}$,  then size of the codebook satisfies:
\begin{equation}\label{AVER2}
\begin{split}
 \frac{M-1}{2} \geq &\underset{0 < \tau_0 < \epsilon}{\sup}\frac{\tau_0 }{\underset{x\in \bar{\mathsf{F}}}{\sup}\beta_{1-\epsilon+ \tau(x)}(x, Q_Y)}.
 \end{split}
 \end{equation}
 This bound is $2\cdot \frac{1}{P_X[\bar{\mathsf{F}}]}$ times larger than the bound with maximal error probability in (\ref{bound20}).
 \end{itemize}
\end{Remarks}

\section{Converse Bound under Maximal Power Constraint in AWGN Channel}\label{C1}

For AWGN channel we have the following specific definitions:
 \begin{itemize}
 \item $\mathsf{A}$ and $\mathsf{B}$ are both $\mathbb{R}^n$.
 \item A vector of length $n$ is usually represented as $\bm{x}$ or $x^n$. The $i$th codeword in the codebook is expressed as $c_i$.
 \item When we use a previous result in the general setting, $x$ stands for a codeword in the general space $\mathsf{A}$.
 \item The condition probability of $Y^n$ when the codeword $\bm{x}$ is sent is $P_{Y^n|X^n=\bm{x}} = \mathcal{N}(\bm{x}, \bm{I}_n)$.
 \item The set $\mathsf{F}_n$ is defined as a $n$-dimensional sphere $\{x^n: \|x^n\|^2 \leq nP\}$
 \item $M_e^*(n,\epsilon, P)$ denotes the maximal number of codewords in a codebook such that each codeword $c_i$ satisfies equal power constraint: $\|c_i\|^2 =  nP$.
 \item $M_m^*(n,\epsilon, P)$ denotes the maximal number of codewords in a codebook such that each codeword $c_i$ satisfies maximal power constraint: $\|c_i\|^2 \leq nP$.
 \item $P$ is a constant unless under the covert constraint, where $P$ is a decreasing function of the blocklength $n$ and written as $P(n)$.

 \end{itemize}

\subsection{Converse Bounds under Maximal Power Constraint And Its Normal Approximation}
In this section, we focus on the converse bound under maximal power constraint and its normal approximation over AWGN channel. The results are applicable in AWGN covert channel if we let $P = P(n)$ with proper $P(n)$. In general, a converse bound is independent of the input distribution and the construction of the code. In the scenario of covert communication, it is assumed that the length of the code as well as the power level is known by the adversary. We prove the converse bound under a maximal power constraint: each codeword $c_i \in X^n$ should satisfy: $\|c_i\|_2^2 \leq nP(n)$.

For the converse bound under maximal probability of error, a general conclusion (Theorem 31 in \cite{Polyanskiy}) under binary hypothesis test is
\begin{equation}\label{Conv1}
M \leq \underset{Q_Y}{\inf} \, \,\underset{x\in \mathsf{F}}{\sup}\frac{1}{\beta_{1-\epsilon}(x, Q_Y)}.
\end{equation}
In the case of maximal power constraint, the distribution $Q_Y$ can be particularized as $\mathcal{N}(\bm{0}, (1+P)\bm{I}_n)$ and the first infimum is removed.
However, we need to find the $\bm{x}$ in $\mathsf{F}_n$ which minimizes $\beta_{1-\epsilon}(\bm{x}, \mathcal{N}(\bm{0}, (1+P)\bm{I}_n))$, which is complicated.
Instead, from a part of the conclusion in Lemma 39 \cite{Polyanskiy}, it shows that
\begin{equation}\label{EandM}
M^*_m(n, \epsilon, P) \leq M^*_e(n+1, \epsilon, P)
\end{equation}
regardless whether $\epsilon$ is an average or maximal probability of error. If a converse bound with blocklength $n+1$ under equal power constraint has been obtained, the above inequality would imply a converse bound with blocklength $n$ under maximal power constraint as follows,
\begin{equation}\label{EandM2}
\begin{split}
M^*_m(n, \epsilon, P)
 \leq &\underset{x^{n+1}\in \mathsf{S}_{n+1}}{\sup}\frac{1}{\beta_{1-\epsilon}(x^{n+1}, Q_{Y^{n+1}})}\\
 = &\frac{1}{\beta_{1-\epsilon}(x^{n+1}, Q_{Y^{n+1}})},
\end{split}
\end{equation}
where $ \mathsf{S}_{n+1}$ is the surface of the $\mathsf{F}_{n+1}$ and $Q_{Y^{n+1}}$ is  $\mathcal{N}(\bm{0}, (1+P)\bm{I}_{n+1})$.
Note that the equality holds in (\ref{EandM2}) because of the sphere symmetry. Thus, the converse bound could be calculated form above inequality by evaluating $P$ as proper $P(n)$ under covert constraint. The details of the evaluation is not included here.

 In \cite{Polyanskiy}, normal approximation of the converse bound under equal power constraint is proved in Theorem 65. However, we could not use it in covert channel directly since the power $P$ is decreasing with $n$ under covert constraint. The following theorem provides normal approximation of the converse bound under equal power constraint with $P = P(n)$.

\begin{Theorem}\label{Conv4}
For AWGN channel with power function $P = P(n)$ (decreasing with $n$),$\epsilon \in (0,1)$ and equal power constraint for each codeword satisfying $\|c_i \|^2 \leq nP(n)$, we have
\begin{equation}\label{Conv2}
\begin{split}
\log &M^*_e(n, \epsilon, P(n)) \\
\leq & nC(n) -\sqrt{nV(n)})Q^{-1}(\epsilon) + \frac{1}{2}\log n + O(1),
\end{split}
\end{equation}
where $C(n) = \frac{1}{2}\log (1 + P(n))$, $V(n) = \frac{P(n)(P(n)+2)}{2(P(n)+1)^2}\log^2e$ and $Q$ function is defined as $Q(x) = \int_{x}^{\infty}\frac{1}{\sqrt{2\pi}}e^{-z^2/2}dz$.
\end{Theorem}
The details of the proof could be found in Appendix \ref{Proof4}.
From (\ref{EandM2}), (\ref{Conv2}) and Taylor's theorem, we have a converse bound under maximal power constraint (maximal probability of error)
\begin{equation}\label{Conv3}
\begin{split}
\log M^*_m(n, \epsilon, P(n)) &\\
\leq nC(n) - &\sqrt{nV(n)})Q^{-1}(\epsilon) + \frac{1}{2}\log n + O(1).
\end{split}
\end{equation}
\hspace{1cm}
\begin{Remarks}
\hspace{0.5cm}
\begin{itemize}
\item
Though the power P is related to n in this case, we can use Berry Esseen
Theorem. The reason will be explained as follows. We use $P = P(n)$ and choose input as
$$\bm{x} = \bm{x_0} =(\underbrace{\sqrt{P(n)}, \sqrt{P(n)}, \cdots, \sqrt{P(n)}}_{m}),$$
then $H_n$ (The formula (\ref{H_n}) in Appendix \ref{Proof4}) is rewritten as $H_m(n)$ . Note that for each m, each coordinate of the $m$-dimensional vector will be identically independent distributed. When we use normal approximation (Berry-Esseen Theorem), any $m$ who satisfies $ m > (\frac{2B_n(P)}{1-\epsilon})^2$ will ensure that $\alpha_m > 1-\epsilon - \frac{2B(P)}{\sqrt{m}} > 0$.  Especially, it is applicable with $m = n$ if $n > (\frac{2B_n(P)}{1-\epsilon})^2$. That is $2B(P)< (1-\epsilon)\sqrt{n}$.
In other words, no matter how small the power of the signal is or how the power decreases with n, we
can always find some N with $2B(P(N)) \leq (1-\epsilon)\sqrt{N}$, and any $n > N$ will be sufficiently large so that
the decoding error probability will be smaller than $\epsilon$.

\item The basis of our proof is that Central Limit Theorem and Berry Esseen Theorem still work provided the power of each coordinate remains constant under any given block length. Actually, when the power $P$ is irrelevant with $n$, the information rate density will approach a fixed Gaussian distribution whose expectation is the capacity and the variance is channel dispersion, both of which are function of $P$. When the power level is decreasing with $n$, the information rate density will approach to Gaussian distributions of different expectations and variances. The key of the proof is that when the power is decreasing with n, the expectations and the variances are closer and closer so that we can approximate them by choosing sufficiently large n.
\end{itemize}
\end{Remarks}

%
\subsection{On the Optimal Output Distribution for Converse Bound}
\begin{Theorem}\label{TightCon}
Under maximal power constraint $P$, the optimal distribution $Q^*_{Y^n}$ of the converse bound which is the solution of the minimax problem
\begin{equation}\label{converseop}
\underset{\mathbb{P}_0}{\inf}\,\underset{\bm{x}\in \mathsf{F}_n}{\sup} \,V_T(\mathbb{P}_1,\mathbb{P}_0)
\end{equation}
where $\mathbb{P}_1 = \mathcal{N}(\bm{x}, \bm{I}_n)$ and $\mathbb{P}_0 = Q_{Y^n}$ and $\mathsf{F}_n$ is the available region of the codewords.
\end{Theorem}
\begin{Proof}
For the converse bound, we have the general result (\ref{Conv1}):
\begin{equation}
M \leq \underset{Q_Y}{\inf}\,\underset{x\in \mathsf{F}}{\sup} \frac{1}{\beta_{1-\epsilon}(x,Q_Y)}
\end{equation}
or
\begin{equation}
1/M  \geq \underset{Q_Y}{\sup}\,\underset{x\in \mathsf{F}}{\inf} \beta_{1-\epsilon}(x,Q_Y)
\end{equation}
where $Q_Y$ is any arbitrary output distribution.
To get the tightest converse bound under given coding scheme, we need to find the $Q_Y$ with which the infimum of $\beta(x)$ with fixed detection probability $\alpha = 1- \epsilon$ obtains its maximum. Each $x$ represents a conditional distribution - a normal distribution with variance $1$ whose mean is at $x$. Each $x$ will induce a curve of ROC (Receiver Operating Characteristic), and $\beta_{1-\epsilon}$ is the x-coordinate corresponding to the point whose y-coordinate is $1-\epsilon$. A typical ROC is plotted in Fig.\ref{Fig3}.

In the case of AWGN channel with maximal power constraint in finite blocklength regime, we have $\mathbb{P}_1 = \mathcal{N}(\bm{x}, \bm{I}_n)$ (the alternative hypothesis) and $\mathbb{P}_0 = Q_{Y^n}$ (the null hypothesis) and there are a lot of such curves, and we want to find a $Q_{Y^n}$ with which the infimum of these $x$-coordinates (false alarm probability $\beta_{1-\epsilon}$) gets its maximum. Now, $1-\epsilon$ is the detection probability and $\beta$ is the false alarm probability which satisfy the following relationship (Theorem 13.1.1 in \cite{E.Lehmann}):
\begin{figure}
\centering
\includegraphics[width=3in]{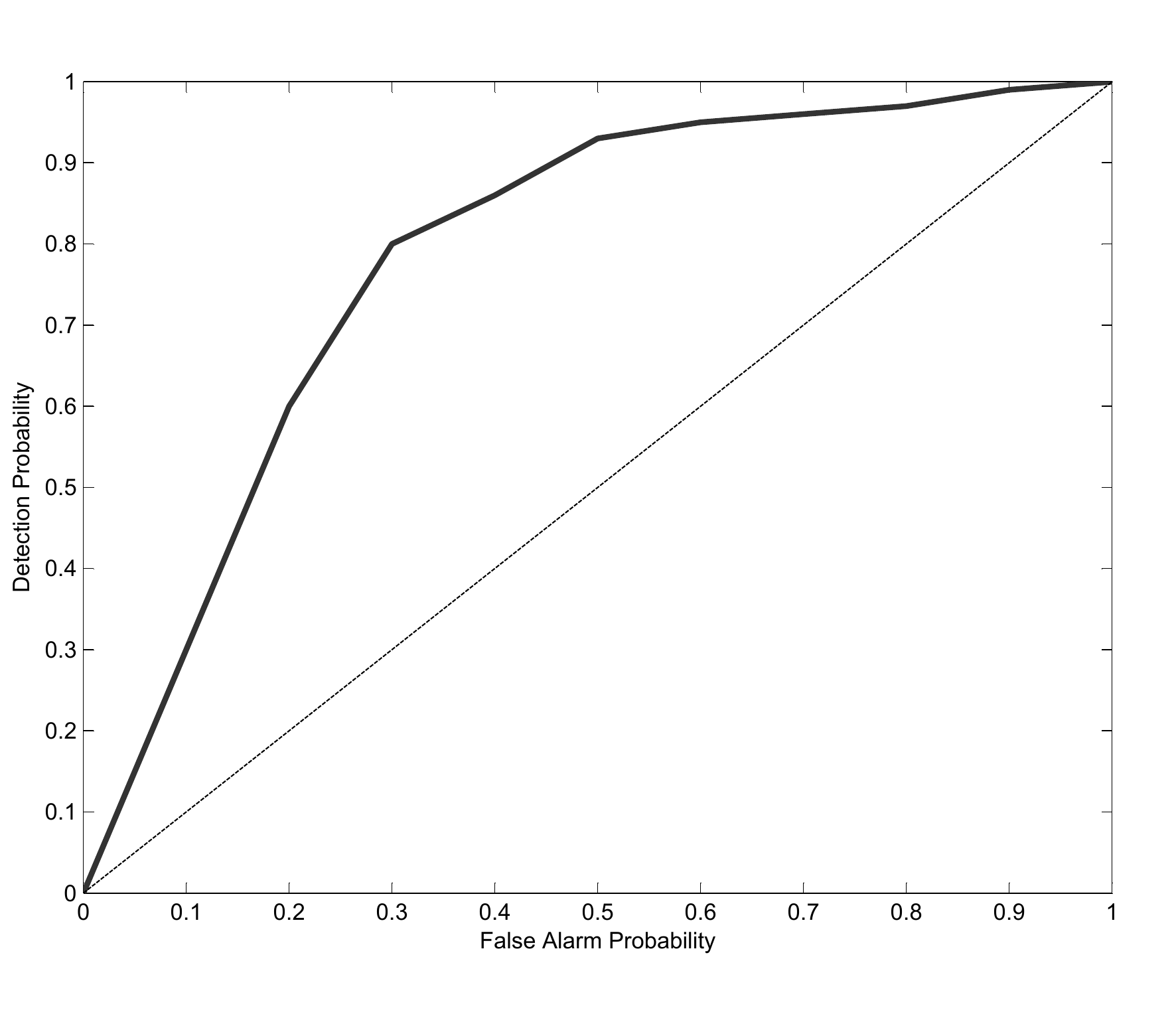}
\caption{A typical curve of ROC}\label{Fig3}
\end{figure}
\begin{equation}\label{equ0}
1 - [\epsilon + \beta_{1-\epsilon}] = V_T(\mathbb{P}_1, \mathbb{P}_0)
\end{equation}
Thus, we have proved the equivalence relation
\begin{equation}
\begin{split}
\frac{1}{M} &\geq \,\underset{\mathbb{P}_0}{\sup}\,\underset{\bm{x}\in \mathsf{F}_n}{\inf} 1-\epsilon -V_T(\mathbb{P}_1,\mathbb{P}_0) \\
 \iff & \frac{1}{M} \geq  1-\epsilon -  \underset{\mathbb{P}_0}{\inf}\,\underset{\bm{x}\in \mathsf{F}_n}{\sup} \,V_T(\mathbb{P}_1,\mathbb{P}_0)\\
\iff & M \leq  \frac{1}{1- \epsilon -\underset{\mathbb{P}_0}{\inf}\,\underset{\bm{x}\in \mathsf{F}_n}{\sup} \,V_T(\mathbb{P}_1,\mathbb{P}_0)}.
\end{split}
\end{equation}
Hence, we want to find some $\mathbb{P}_0 = Q_{Y^n}$, the total variation distance between it and the conditional distribution $\mathcal{N}(\bm{x},\bm{I}_n)$ with $\bm{x}\in \mathsf{F}_n$ is uniformly minimized.
\end{Proof}
In general, the optimization problem (\ref{converseop}) is not easy to solve. To get a resolvable problem, we use Pinsker's inequality:
\begin{equation}
V_T(\mathbb{P}_1,\mathbb{P}_0) \leq \sqrt{\frac{1}{2}D(\mathbb{P}_1\|\mathbb{P}_0)}.
\end{equation}
Thus, a more manageable upper bound  on $M$ can be formulated as
\begin{equation}
1-\epsilon - \frac{1}{M} \leq \,\underset{\mathbb{P}_0}{\inf}\,\underset{\bm{x}\in \mathsf{F}_n}{\sup} \,\sqrt{\frac{1}{2} \,D(\mathbb{P}_1\|\mathbb{P}_0)}
\end{equation}
or
\begin{equation}\label{minmax1}
1-\epsilon - \frac{1}{M} \leq  \sqrt{\frac{1}{2} \underset{\mathbb{P}_0}{\inf}\,\underset{\bm{x}\in \mathsf{F}_n}{\sup} \,D(\mathbb{P}_1\|\mathbb{P}_0)}
\end{equation}
where $\mathbb{P}_1 = \mathcal{N}(\bm{x}, \bm{I}_n)$  and $\mathbb{P}_0 = Q_{Y^n}.$
Now all these $\bm{x}\in \mathsf{F}^n$ have an induced channel which is a vector Gaussian channel;
\begin{equation}\label{vectorchannel}
\bm{y} = \bm{x} + \bm{z}
\end{equation}
where $\bm{x}\in \mathbb{R}^n$ is independent of $\bm{z} \in \mathbb{R}^n$ and $\bm{z} \sim \mathcal{N}(\bm{0}, \bm{I}_n)$.
In our case, the maximal power constraint $\bm{x}\in \mathsf{F}^n$ which is actually the constraint that $\bm{x} \in \mathfrak{B}^n_0(\sqrt{nP})$ where $\mathfrak{B}^n_0(\sqrt{nP})$ is the n-dimensional sphere centered at $0$ of radius $\sqrt{nP}$. It is equivalent to a peak power constraint on the input vector $\bm{x}$.  From the duality of channel coding and universal source coding (Chapter 13 in \cite{Cover}), the solution of $\mathbb{P}_0$ in (\ref{minmax1}) is the induced output distribution of the capacity-achieving distribution of this channel (\ref{vectorchannel}). The solution to this problem has been investigated in \cite{Borzoo} and \cite{Alex} and it has been shown that the support of
the capacity-achieving distribution is a finite set of hyper-spheres
with mutual independent phases and amplitude in the spherical
domain. A uniform distribution on a single sphere is optimal as $\frac{R}{\sqrt{n}} \rightarrow 0$ has been shown in \cite{Borzoo}. In general case when $R=\sqrt{n P}$ and $R/\sqrt{n} = \sqrt{P}$ is a constant, the uniform distribution of the surface of the sphere is not optimal.
A tight converse bound is from (\ref{Conv1}). We have to resolve the underlying minimax problem to get the optimal $Q_{Y^n}$ to get the tightest converse bound.

If we further impose the covert constraint, we will have $P(n) \rightarrow 0$, as $n$ goes to $\infty$. That means for the covert channel, a converse bound using (\ref{minmax1}) with codewords chosen on a surface of a sphere is reasonable. Even so, the output distribution cannot be Gaussian for finite $n$. In our case and the converse bound with maximal power constraint (218) in \cite{Polyanskiy}, the choice of the auxiliary output distributions as Gaussian is for computation convenience. This fact implies that our converse bound and the converse bound (218) in \cite{Polyanskiy} are not the tightest bounds. However, evaluation of the tightest bound under such optimal APSK (amplitude and phase shift keying) support set is beyond the scope of this paper, and will be addressed in our future works.
\section{Achievability Bound under Maximal Power Constraint in AWGN Channel}
In this section, the general results in Section \ref{A1} will be applied to obtain achiavability bounds over AWGN channel. These result can be further applied in covert channel. The definitions and notions are almost the same as Section \ref{C1}. Below are some new items needed to derive our achievability bound.
\begin{itemize}
\item $\mu$ is a parameter to constrain the candidates of codewords, which may depend on $n$.
\item For each $n$,  $\bar{\mathsf{F}}_n \triangleq \{x^n: \mu^2 \cdot nP\leq \|\bm{x}\|_2^2 \leq nP\}$.
\end{itemize}
 As in last section, $P$ will be regarded as a constant unless under covert constraint, where it is written as $P(n)$.
\subsection{The Achievability Bound over AWGN Channel under Maximal Power Constraint}
We will now apply Corollary \ref{AcCorollary1} in AWGN channel with blocklength $n$ under maximal power constraint.
The input distribution is $P_{X^n} = \mathcal{N}(\bm{0}, \mu P\bm{I}_n)$ and the set $\bar{\mathsf{F}}$ will be $\bar{\mathsf{F}}_n$. Based on this subset, the truncated distribution is $\bar{P}_{X^n}$, We use a constant $0 <\tau < \epsilon$ instead of $\tau_0$ for convenience and the bound (\ref{bound2}) can be rewritten as
\begin{equation}\label{AC3}
M \geq \underset{0< \tau < \epsilon}{\sup} \frac{\tau }{\sup_{x^n \in \bar{\mathsf{F}}_n}\beta_{1-\epsilon+ \tau}(x^n, P_{Y^n})}.
\end{equation}
The construction of the code is the the same as Lemma \ref{Le1}. Note that $P_{Y^n}$ is the induced distribution of $\bar{P}_{X^n}$.
\subsection{The Computation of Achievability Bound}\label{Calculation}
As $P_{Y^n}$ is the induced output distribution of a truncated $n$-dimensional Gaussian distribution, it brings much difficulty to the evaluation of (\ref{AC3}). For computation convenience, we substitute $Q_{Y^n} = \mathcal{N}(\bm{0}, (1 + \mu P)\bm{I}_n)$ for $P_{Y^n}$. From Neyman-Pearson Lemma, we have
\begin{eqnarray}\label{AC4}
M \geq &\underset{0< \tau < \epsilon}{\sup} \frac{\tau }{\sup_{x^n \in \bar{\mathsf{F}}_n}\beta_{1-\epsilon+ \tau}(x^n, P_{Y^n})} \notag\\ \geq &\underset{0< \tau < \epsilon}{\sup} \frac{\tau }{\sup_{x^n \in \bar{\mathsf{F}}_n}\beta_{1-\epsilon+ \tau}(x^n, Q_{Y^n})}.
\end{eqnarray}

\begin{figure}
\centering
\includegraphics[width=3in]{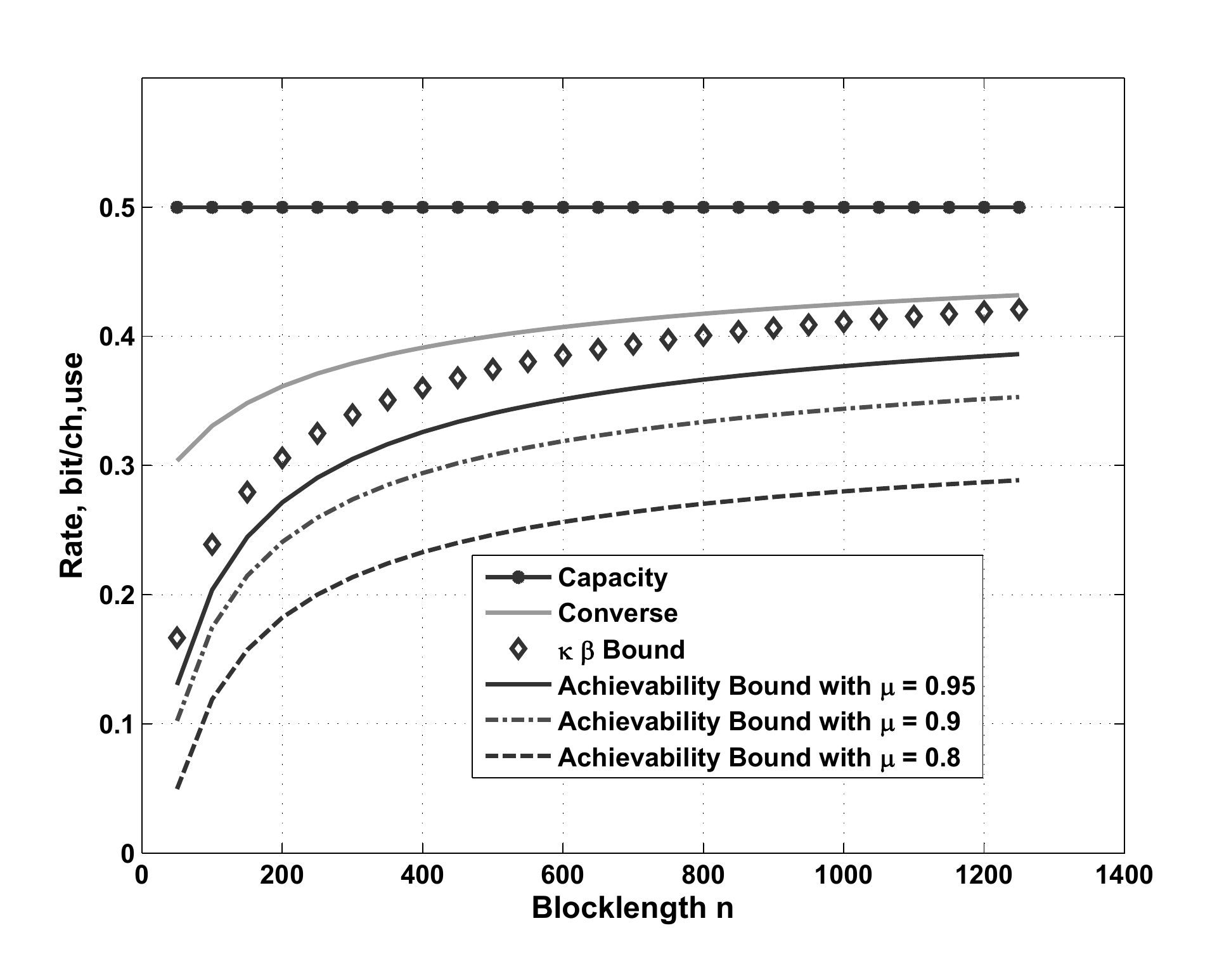}
\caption{Comparison of the achievability bounds, where $SNR = 0 \,dB$ and $\epsilon = 10^{-3}$.}\label{Fig11}
\end{figure}

  Note that $Q_{Y^n}$ is precisely the induced output distribution of $P_{X^n}$.
The numerical evaluation will be direct computation of (\ref{AC4}).
Let us explain how to compute the above bound:
\begin{enumerate}
\item On the surface of $n$-dimensional sphere with radius $\sqrt{nR}$, $\beta_{1-\epsilon+ \tau}(x^n, Q_{Y^n})$ will be all the same due to sphere symmetry. We can choose a particular $x^n = [\sqrt{R}, \sqrt{R}, \cdots, \sqrt{R}]$ to compute $\beta_{1-\epsilon+ \tau}(x^n, Q_{Y^n})$, then search the largest $\beta_{1-\epsilon+ \tau}(x^n, Q_{Y^n})$ on $\mu^2\cdot P(n)\leq R \leq P(n)$ for each fixed $\tau \in (0, \epsilon)$.
\item  The computation of $\beta_{1-\epsilon+ \tau}(x^n, Q_{Y^n})$ is from Theorem 40 in \cite{Polyanskiy} except that $G_n$ and $H_n$ are substituted by $G_n(R)$  and $H_n(R)$, respectively. Please refer to the details in \cite{Yu2}.
\item Search a suitable $\tau$ in the interval $0 < \tau < \epsilon$ to find $\underset{0<\tau <\epsilon}{\sup}  \frac{\tau }{\sup_{x^n \in \bar{\mathsf{F}}_n}\beta_{1-\epsilon+ \tau}(x^n, Q_{Y^n})}$.
\end{enumerate}
In Fig.\ref{Fig11}, we compare our achievability bound with the $\kappa \beta$ bounds in \cite{Polyanskiy} when the parameter $\mu$ has different values. It is obvious from the construction that $\kappa\beta$ bound is almost the special case of $\mu = 1$. When $\mu$ tends to $1$, the achievability bound will approach to $\kappa\beta$ bound. In fact, the constraint from the inside of the sphere is to ensure that $\sup_{x^n \in \mathsf{F}_n}\beta_{1-\epsilon + \tau}$ is comparable with $\beta_{1-\epsilon + \tau}(x^n, Q_{Y^n})$ with the specific $x^n = [\sqrt{P}, \cdots, \sqrt{P}]$. The converse bound in Fig.\ref{Fig11} is from (\ref{EandM2}). Thus, it is almost the same as the converse bound in \cite{Polyanskiy} with equal power constraint.

As we have seen, replacing $P_{Y^n}$  by $Q_{Y^n} = \mathcal{N}(\bm{0}, (1+P(n)\bm{I}_n))$ will simplify the evaluation. In fact, the benefit of this substitution is more than that. In next section, we will continue to utilize the advantage of the simplification to obtain normal approximation of the bound. It is no doubt that the substitution will impair the tightness of the obtained bound. Nevertheless, the impairment by the replacement is insignificant when the blocklength is moderately large and $\mu$ is properly chosen. This topic will be discussed later in Section \ref{Discussion}.

Though our achievability bound is not tighter than the $\kappa\beta$ bound under equal power constraint, it is meaningful and necessary for the analysis on asymptotic throughput of covert communication. The reason is that our coding scheme is Gaussian random coding with codewords not placed on the surface of a sphere as assumed in \cite{Polyanskiy} and the known results in the liternate can not be applied directly. The analysis of throughput on Gaussian random coding relies on integrating the following three ingredients: finite blocklength, random coding scheme and maximal power constraint.

\subsection{Normal Approximation of The Achievability Bound over AWGN Channel}\label{NorAC}
In this section, we investigated normal approximation of the achievability bound of Gaussian random coding under maximal power constraint $P= P(n)$. Generally speaking, the framework of the proof on achievability bound here is similar as the proof of Theorem 67 in \cite{Polyanskiy}. However, there are several differences.
 \begin{itemize}
 \item Firstly, in the generation of the codebook, the input distribution is determined as zero-mean Gaussian distribution with a specific variance $\mu P(n)$. The dependence on both $\mu$ and $P(n)$ stems from maximal power constraint (controlling the TVD at the adversary under covert constraint).
 \item Secondly, the codewords are drawn from an specific set $\bar{\mathsf{F}}_n$ which is a subset of a $n$-dimension sphere and varies with $n$.
 \item Thirdly, as in the proof of the converse bound, the power $P$ is decreasing with $n$, so that we should be very cautious when dealing with normal approximation.
 \end{itemize}

The base of the normal approximation is formula (\ref{AC4}):
\begin{equation}\notag
 M \geq \underset{0< \tau < \epsilon}{\sup} \frac{\tau }{\sup_{x^n \in \bar{\mathsf{F}}_n}\beta_{1-\epsilon+ \tau}(x^n, Q_{Y^n})}.
\end{equation}

\begin{Theorem}\label{AC1}
For AWGN channel with noise $\mathcal{N}(0, 1)$ and any $0 <\epsilon < 1$, there exists an $(n, M, \epsilon)$ code (maximal probability of error) chosen from a set $\bar{\mathsf{F}}_n$ of codewords whose coordinates are i.i.d $\sim \mathcal{N}(0, \mu P(n))$ where $0 <\mu < 1$ and also satisfy
 \begin{enumerate}
 \item $ \mu^2 \cdot nP(n)\leq \|\bm{x}\|_2^2 \leq nP(n)$,
 \item  $\tau_0 \leq \tau_n(R) \leq \frac{n}{n+1}\epsilon$.
 \end{enumerate}
Let
\begin{equation*}
\begin{split}
&\bm{x} = [\sqrt{R}, \cdots, \sqrt{R}], \, \  \,C_{\mu}(n) = \frac{1}{2}\log(1 + \mu P(n)),\\
&\tau^{\mu}_n(R) =\frac{B_{\mu}(P,R)}{\sqrt{n}},\, \  \, B_{\mu}(P, R) = \frac{6T_{\mu}(P, R)}{\hat{V}_{\mu}(P, R)^{3/2}},\\
&T_{\mu}(P, R) = \mathbb{E}\left[|\frac{\log e}{2(1+\mu P)}[\mu P + 2\sqrt{R}Z_i -\mu PZ_i^2]|^3\right],\\
&\hat{V}_{\mu}(P, R) = \left(\frac{\log e}{2(1+P)}\right)^2(4R+ 2P^2) = V(n) \cdot \left(\frac{2R+ P^2}{2P+ P^2}\right),
\end{split}
\end{equation*}
where $Z_i$'s are i.i.d standard normal,
then we have (maximal probability of error)\footnote{Due to the revision in (\ref{bound2}) of Corollary \ref{AcCorollary1} based on \cite{Yu2}, the term $P_X[\mathsf{F}]$ is abandoned in the following expressions of the bound in this work.}
\begin{equation}\label{Achi3}
\begin{split}
\log M^*_m(n, \epsilon, P(n)) \geq   \underset{0 < \tau_0 < \epsilon}{\sup}& \{ nC_{\mu}(n) +  \frac{n(R^* -\mu P(n))\log e}{2(1 + \mu P(n))}\\
+ \sqrt{n\hat{V}_{\mu}(P(n),R^*)}Q^{-1}&\left(1 - \epsilon + \frac{2B_{\mu}(P(n),R^*)}{\sqrt{n}}\right) \\
 + \log\tau_0 + \frac{1}{2}\log n
- \log& \left[\frac{2\log 2}{\sqrt{2\pi \hat{V}_{\mu}(P, R^*)}}+ 4B_{\mu}(P, R^*)\right]\}.
\end{split}
\end{equation}
The quantity $R^*$ satisfies $ x^n_0 = [\sqrt{R^*}, \cdots, \sqrt{R^*}] \in \bar{\mathsf{F}}_n$  and maximizes
\begin{equation}\label{logM1}
\begin{split}
&nC_{\mu}(n) +  \frac{n(R -\mu P(n))\log e}{2(1 + \mu P(n))} + \frac{1}{2} \log n + \log\tau_0  \\
 + & \sqrt{n\hat{V}_{\mu}(P(n),R)}Q^{-1}\left(1 - \epsilon + \frac{2B_{\mu}(P(n),R)}{\sqrt{n}}\right)\\
 - &\log \left[\frac{2\log 2}{\sqrt{2\pi \hat{V}_{\mu}(P, R)}}+ 4B_{\mu}(P, R)\right].
\end{split}
\end{equation}
\end{Theorem}
\noindent The details of the proof could be found in Appendix \ref{Proof6}

\begin{Remarks}
\hspace{0.5cm}
\begin{itemize}
\item The condition $\tau_0 \leq \tau_n(R) \leq \frac{n}{n+1}\epsilon$ is necessary for normal approximation because the quantity $\beta_{1-\epsilon+ \tau^{\mu}_n(R)}$ with different $R$ will lead to different offsets in normal approximation.
\item If we could prove that $R = P(n)$ satisfies $\tau_0 \leq \tau^{\mu}_n(R) < \epsilon$ and maximizes $\beta_{1-\epsilon+ \tau_n^{\mu}(R)}$ or
\begin{small}
\begin{equation}\label{maxim}
\begin{split}
&\frac{n(R -\mu P(n))\log e}{2(1 + \mu P(n))} - \log \left[\frac{2\log 2}{\sqrt{2\pi \hat{V}_{\mu}(P, R)}}+ 4B_{\mu}(P, R)\right]\\
&+ \sqrt{n\hat{V}_{\mu}(P(n),R)}\times Q^{-1}
\left(1 - \epsilon + \frac{2B_{\mu}(P(n),R)}{\sqrt{n}}\right),\\
\end{split}
\end{equation}
\end{small}
 From the expression of (\ref{maxim}), as a function of $R$ with fixed $n$, the derivative of it is positive when $n$ is sufficiently large and there exists some $\tau_0$ that $\epsilon > \tau_n(R) > \tau_0$ holds. Hence, we have
 \begin{equation}\label{achibound1}
    \begin{split}
     \log M^*_m(n, \epsilon, P(n)) \geq & nC_{\mu}(n) -\sqrt{nV(n)})Q^{-1}(\epsilon) \\
     + \frac{1}{2}\log n  + &\log\tau_0 + O(1)
     \end{split}
    \end{equation}
 holds for some $\tau_0$.
  The above claim holds when $n$ is sufficiently large.
\item The right-hand side of the constraint $\tau_0 \leq \tau^{\mu}_n(R) \leq \frac{n}{n+1}\epsilon$ is to ensure that the set $R_n$ (The definition of $R_n$ can be found in the proof) is compact. $\tau_0 \leq \tau_n^{\mu}(R) < \epsilon$ is sufficient for the existence of $\bar{\mathsf{F}}_n$ and the upper bound (\ref{maxim}) of $\beta_{1-\epsilon+ \tau^{\mu}_n(R)}$.
\item The region of the candidates for the codewords is constrained from both the outside of the sphere and the inside of the sphere. The constraint $\|\bm{x}\|_2^2 \leq nP(n)$ is to satisfy the maximal power constraint induced from covert constraint. The constraint from inside of the sphere is necessary because it will lead to an achievability bound comparable with respect to \cite{Polyanskiy}. We will explain it in the next section.
\item The parameter $\mu$ satisfies $\mu\in (0, 1]$. If $\mu$ is close to $0$, the set $\mathsf{F}_n$ is almost the whole space and the codewords are almost i.i.d Gaussian distributed. The bound will become trivial. Nevertheless, $\mu$ can be slightly less than $1$ and the codewords are still behaving like Gaussian codewords due to sphere hardening effect for large $n$ \cite{Hamkins}. The utilization of sphere hardening effect is important for controlling the TVD at the adversary. The details about choosing $\mu$ will be discussed later.
\end{itemize}
\end{Remarks}

\subsection{On the Optimal Input Distribution for Achievability Bound}

\begin{Theorem}\label{TightAc}
Let the codewords be generated from a distribution $P_{X^n}$ whose support is a subset $\mathsf{F}^n$ of space $\mathbb{R}^n$, the optimal distribution $P^*_{X^n}$ for the achievability bound under the coding scheme is the solution of the optimization problem
\begin{equation}\label{Achiop}
\underset{P_{X^n}: supp(P_{X^n}) \subseteq \mathsf{F}^n}{\sup} \,\underset{\bm{x}\in \mathsf{F}^n}{\inf} \, V_T(\mathbb{P}_1, \mathbb{P}_0)
\end{equation}
where $supp$ denotes the support of $P_{X^n}$, $\mathbb{P}_0 = P_{Y^n}$ is the output distribution induced by $P_{X^n}$ and $\mathbb{P}_1$ is the conditional distribution $\mathcal{N}(\bm{x},\bm{I}_n)$ with $\bm{x}\in \mathsf{F}^n$.
\end{Theorem}
\begin{Proof}
In the general setting,  the achievability bound follows Corollary \ref{AcCorollary1},
\begin{equation}\label{bound21}
M \geq \underset{0 < \tau < \epsilon}{\sup}\frac{\tau}{\underset{x\in \bar{\mathsf{F}}}{\sup}\beta_{1-\epsilon+ \tau}(x, P_Y)}.
\end{equation}
where $P_Y$ is the output distribution induced by the codewords whose support is $\bar{\mathsf{F}}\subset \mathsf{A}$. In order to get the tightest bound, $\bar{\mathsf{F}}$ should be related to $\tau$ and we denote it as $\mathsf{F}_\tau$.
The tightest achievability bound for this random coding scheme can be obtained as
\begin{equation}
M \geq \underset{0 < \tau < \epsilon}{\sup}\underset{P^{\tau}_X: supp(P^{\tau}_X) \subseteq \mathsf{F}_{\tau}}{\sup}\frac{\tau}{\underset{x\in\mathsf{F}_\tau}{\sup}\beta_{1-\epsilon+ \tau}(x, P_Y)}
\end{equation}
From the above inequality and the relationship between $\beta$ and TVD (\ref{equ0}), we have
\begin{equation}
\begin{split}
\frac{1}{M} &\leq \underset{\tau}{\inf}\, \underset{P^{\tau}_X: supp(P^{\tau}_X) \subseteq \mathsf{F}_{\tau}}{\inf} \,\underset{x\in \mathsf{F}_{\tau}}{\sup} \frac{1}{\tau}\beta_{1-\epsilon + \tau}(x,P_Y)\\
\iff &\frac{1}{M} \leq \underset{\tau}{\inf}\, \underset{P^{\tau}_X: supp(P^{\tau}_X) \subseteq \mathsf{F}_{\tau}}{\inf} \,\underset{x\in \mathsf{F}_{\tau}}{\sup} \frac{1}{\tau}[1 - \epsilon + \tau - V_T(\mathbb{P}_1, \mathbb{P}_0)]\\
\iff & \frac{1}{M} \leq \underset{\tau}{\inf}\,\frac{1}{\tau}\left[1 - \epsilon + \tau -\underset{P^{\tau}_X: supp(P^{\tau}_X) \subseteq \mathsf{F}_{\tau}}{\sup} \,\underset{x\in \mathsf{F}_{\tau}}{\inf} V_T(\mathbb{P}_1, \mathbb{P}_0)\right].
\end{split}
\end{equation}
where $\mathbb{P}_1$ is the induced distribution of the particular $x$ and $\mathbb{P}_0$ is the induced distribution of all the codewords.
In the case of AWGN channel in finite blocklength regime, the inner optimization problem is rewritten as
\begin{equation}\label{op1}
\underset{P^{\tau}_{X^n}: supp(P^{\tau}_{X^n}) \subseteq \bar{\mathsf{F}}_{\tau}^n}{\sup} \,\underset{
\bm{x}\in \bar{\mathsf{F}}_{\tau}^n}{\inf} V_T(\mathbb{P}_1, \mathbb{P}_0)
\end{equation}
where $\mathbb{P}_1 = \mathcal{N}(\bm{x}, \bm{I}_n)$ and $\mathbb{P}_0$ is the induced output distribution of $P^{\tau}_{X^n}$.
 Hence, we want to find some $\mathbb{P}_0$, the infimum of TVD between it and the conditional distributions $\mathcal{N}(\bm{x},\bm{I}_n)$ with $\bm{x}\in \bar{\mathsf{F}}_{\tau}^n$ is maximized. Hence, the inner optimal distribution for the achievability bound is the solution of (\ref{op1}).
The tightest bound should be obtained by choosing the optimal $\tau$ which maximizes it. Now if the set $\mathsf{F}$ is fixed, then the inner optimization problem is irrelevant with $\tau$, hence is (\ref{Achiop}).
\end{Proof}
 In this paper, only Gaussian input distribution with additional max power constraint on the codewords are considered. As shown in the previous theorem, the optimal input distribution for attaining a tight achievability throughput bound is not necessarily Gaussian.

\begin{figure}
\centering
\includegraphics[width=3in]{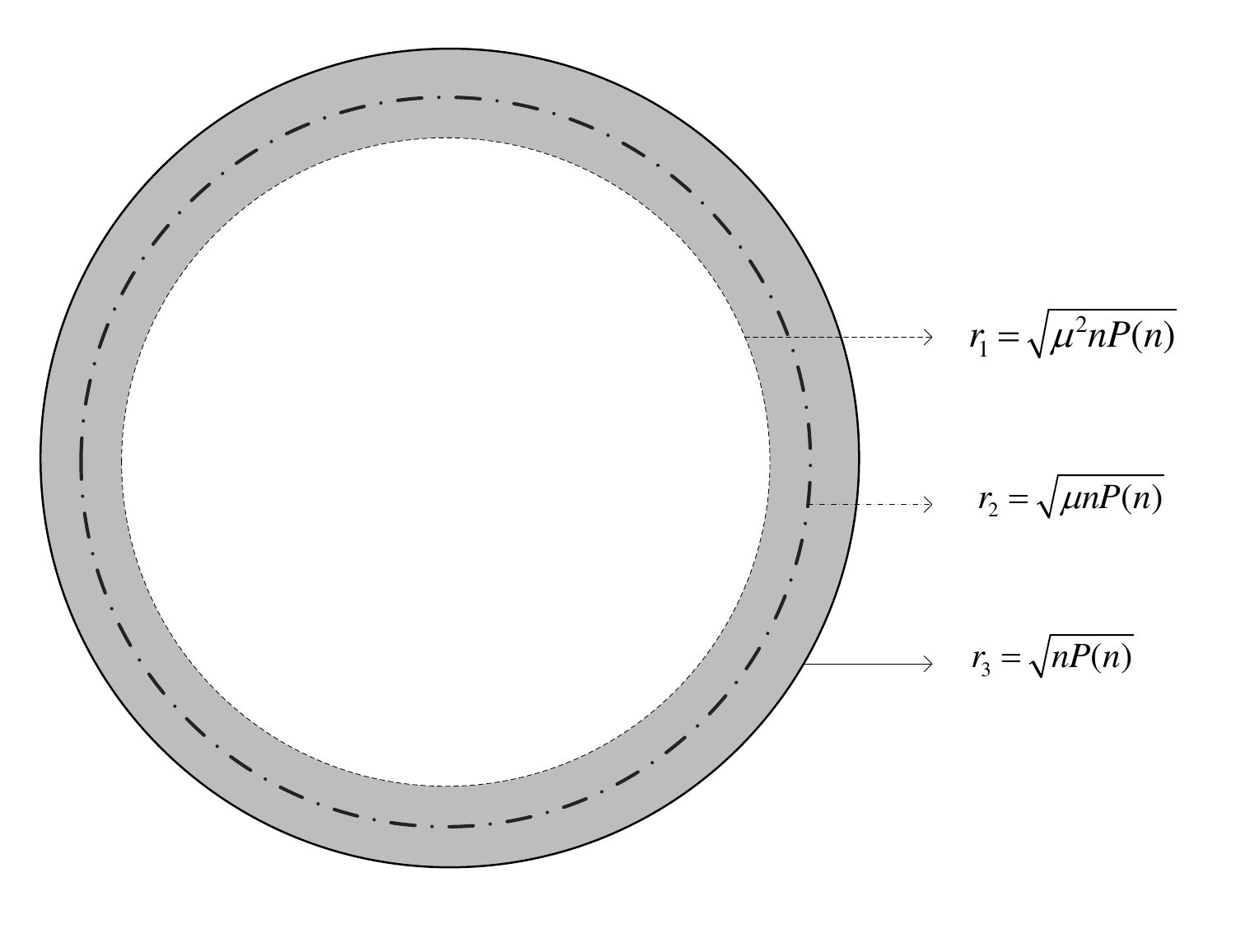}
\caption{The candidates of codewords lies in a subset of n-dimensional sphere: $\mathsf{F}_n \triangleq \{x^n: \mu^2 \cdot nP(n)\leq \|x^n\|_2^2 \leq nP(n) \}$.}\label{Fig2}
\end{figure}

\subsection{Further Discussions on the Bounds}\label{Discussion}

There is no doubt that our achievability bound is based on the achievability ound of Theorem 21 in \cite{Polyanskiy} and it looks like the $\kappa\beta$ bound. Nevertheless, there are several differences between our achievability bound and the existing ones. Firstly, the codebook of $\kappa\beta$ bound is deterministic because there is no distribution on $X$. Though it has a extension (127) with input distribution $Q_X$, the construction of the codebook is irrelevant with $Q_X$. In the literature such as \cite{Polyanskiy},\cite{Shannon} and \cite{Tan}, the codewords in AWGN channel are subjected to equal power constraint, i.e. all the codewords are on the surface of n-dimensional sphere whose radius is $r_3 = \sqrt{nP(n)}$ with $P(n) = P$. In this work, the codewords are scattered across the shaded area in Fig.\ref{Fig2} as maximal power constraint and random coding are adopted in the finite blocklength regime. Secondly, for moderate blocklength $n$ and properly chosen $\mu$, the codewords can be regarded as Gaussian codewords in some extent. This is due to sphere hardening effect \cite{Hamkins}. In fact, if a Gaussian codebook is adopted, i.e., each coordinate of these codewords is independently drawn from $\mathcal{N}(\bm{0}, \mu P(n)\bm{I}_n)$, then most of them shall fall into the shaded region in Fig.\ref{Fig2} for $\mu = 0.8$ and $n \geq 400$. That is, the impact of truncation is insignificant with moderate blocklength $n$ and proper chosen $\mu$. This also implies that the replacement of $P_{Y^n}$ for $Q_{Y^n}$ has little impairment on the bound since if $\bar{P}_{X^n}$ is close to $P_{X^n}$, $Q_{Y^n}$ will be also close to $P_{Y^n}$. Consequently, the computation method in Section \ref{Calculation} is adequately accurate for most applications. Moreover, the normal approximation in Section \ref{NorAC} provides a convincing estimation of the achievability bound of the throughput.

From Theorem \ref{TightCon} and Theorem \ref{TightAc}, the optimal output distribution of the tightest converse bound under equal power constraint for finite $n$ should be the solution of the minimax problem (\ref{converseop}) with $\mathsf{F}$ being the surface of the sphere. The optimal input distribution of the tightest achievability bound under equal power constraint for finite $n$ should be the solution of the minimax problem (\ref{Achiop}) with $\mathsf{F}$ being the surface of the sphere. However, the output distributions in evaluation of both bounds are the same (zero-mean Gaussian distribution).
In either case, we have reasons to question how far these bounds are from the ones to be found using the minimax problem of TVD, which will be presented in our future works.

\section{Application in Covert Communication and Numerical Results}

In this section, the application of previous results is introduced. Before we consider the attainable throughput of covert communication with finite blocklength, it should be clarified that both the achievability bound and converse bound are meaningful only when the covert constraint is satisfied. The main concern includes two aspects: (1) With given blocklength $n$, how to choose proper parameters of the coding scheme to meet the covert constraint of TVD? (2) How could these achievability bounds be applied to covert communications?

\subsection{TVD Requirement at the Adversary}

The first concern is related to the selection of the parameters $P(n)$ and $\mu$. As we have introduced in Section II, the covert constraint is in the form of an upper bound $\delta$ on the TVD between the distributions of eavesdropped signals at an adversary with and without presence of active and legitimate communication, respectively. As TVD is a normalized metric, $\delta$ near $0$ is usually adopted. From now on, we focus on the induced output distribution of the codes at the adversary.
 Note that our codewords are selected from a subset of $\mathbb{R}^n$, there is a distinction between the input distribution and Gaussian distribution when the blocklength is small.

 Recall the process of generating the codebook: each coordinate of the candidates is generated from i.i.d Gaussian distribution $\mathcal{N}(0, \mu P(n))$ and then each codeword is selected within the region where the radius is between $\sqrt{\mu^2nP(n)}$ and $\sqrt{nP(n)}$ as shown in Figure \ref{Fig2}. The distribution $\bar{P}_{X^n}$ of the codewords is a truncated Gaussian distribution whose density function is
\begin{equation}
\footnotesize
\bm{f}(\bm{x}) = \left\{
\begin{split}
&\frac{1}{\Delta}\frac{1}{(2\pi \mu P(n))^{k/2}}e^{-\frac{\|\bm{x}\|^2}{2\mu P(n)}},    \sqrt{\mu^2nP(n)} \leq \hspace{-0.04in}\|\bm{x}\|\hspace{-0.04in}\leq \hspace{-0.04in}\sqrt{nP(n)}\\
&0,    \ \  \  \  \  \  \  \  \ \  \  \  \  \  \  \  \  \  \  \  \  \  \  \  \  \ \ \  \  \  \  \  \  \  \  \,otherwise,
\end{split}
\right.
\end{equation}
where $\Delta$ is the normalized coefficient
 \begin{equation}
 \Delta = E[1_{\{\bm{x} \in \mathfrak{B}^n_0(\sqrt{nP(n)})\backslash \mathfrak{B}^n_0(\sqrt{\mu^2nP(n)})\}}].
 \end{equation}
 As the distribution of the candidates $P_{X^n}$ has density function
\begin{equation}
\bm{g}(\bm{x}) = \frac{1}{(2\pi \mu P(n))^{k/2}}e^{-\frac{\|\bm{x}\|^2}{2\mu P(n)}},
\end{equation}
the pdf of $\|\bm{x}\|$ is expressed as (refer to \cite{Hamkins})
\begin{equation}\label{matu}
h(r) = \frac{2r^{n-1}e^{ -\frac{r^2}{2\mu P(n)}}}{\Gamma(n/2)(2\mu P(n))^{n/2}}.
\end{equation}

Let $\mathbb{P}_0$ be the n-dimensional noise distribution $\mathcal{N}(\bm{0}, \bm{I}_n)$, $\mathbb{P}_1$ be the output distribution induced by the n-dimensional Gaussian distribution $\mathcal{N}(\bm{0}, \mu P(n)\bm{I}_n)$ and let $\bar{\mathbb{P}}_1$ be the output distribution of the truncated Gaussian distribution $\bar{P}_{X^n}$. From above analysis, TVD at the adversary is written as
\begin{equation}\label{TVD1}
V_T(\bar{\mathbb{P}}_1, \mathbb{P}_0)
\end{equation}
and the power level should be chosen so that $ V_T(\bar{\mathbb{P}}_1, \mathbb{P}_0) \leq \delta$.
It is difficult to get an analytic formula of (\ref{TVD1}). We use the following bounds of TVD at the adversary.
\begin{Triangular Inequality Bound}
TVD is a distance and satisfies the triangle inequality \cite{Tsybakov}:
\begin{equation}\label{TVD2}
\begin{split}
|V_T(\mathbb{P}_1, \mathbb{P}_0) - V_T(\bar{\mathbb{P}}_1, \mathbb{P}_1)|&\leq V_T(\bar{\mathbb{P}}_1, \mathbb{P}_0) \\ &\leq V_T(\mathbb{P}_1, \mathbb{P}_0) + V_T(\bar{\mathbb{P}}_1, \mathbb{P}_1).
\end{split}
\end{equation}
\end{Triangular Inequality Bound}

\begin{Data Processing Inequality Bound}
\begin{equation}
V_T(\bar{\mathbb{P}}_1, \mathbb{P}_1)\leq V_T(\bar{P}_{X^n}, P_{X^n}).
\end{equation}
\end{Data Processing Inequality Bound}
Nota that both $\bar{P}_{X^n}$ and $P_{X^n}$ are absolutely continuous respect to Lebesgue measure, and have their corresponding density functions. The last inequality is an application of the following theorem \cite{Rudin} by letting $f$ be the difference of the density functions of $\bar{P}_{X^n}$ and $P_{X^n}$ and $g$ be the density function of i.i.d n-dimensional Gaussian noise.
\begin{Theorem}
Suppose $f, g \in L^1(\mathbb{R}^n)$. Then for a.e. $x\in \mathbb{R}^n$, the convolution $(f*g)(x)$ exists, $f*g \in L^1(\mathbb{R}^n)$, and
\begin{equation}
\|f*g\|_1 \leq \|f\|_1\cdot\|g\|_1.
\end{equation}
\end{Theorem}
\noindent From the above bounds, we have an upper bound of $ V_T(\bar{\mathbb{P}}_1, \mathbb{P}_0)$:
\begin{equation}\label{Upper}
 V_T(\bar{\mathbb{P}}_1, \mathbb{P}_0) \leq V_T(\mathbb{P}_1, \mathbb{P}_0) + V_T(\bar{P}_{X^n}, P_{X^n}).
\end{equation}
\noindent Now the quantity $V_T(\bar{P}_{X^n}, P_{X^n})$ could be computed as follows
\begin{equation}\label{trunction}
V_T(\bar{P}_{X^n}, P_{X^n}) = \frac{1}{2}\underset{\mathcal{R}^n}{\int}|\bm{f}(\bm{x}) - \bm{g}(\bm{x})|d\bm{x} = 1- \Delta.
\end{equation}
From (\ref{matu}), after integrate $r$ from $0$ to $\sqrt{nP(n)}$, we have
\begin{equation}
\Delta =  \frac{\gamma(n/2, n/2\mu)- \gamma(n/2, n\mu/2)}{\Gamma(n/2)}
\end{equation}
where $\gamma(a,z)$ is incomplete gamma function defined as follows
 \begin{equation}
 \gamma(a, z)= \int_0^{z}e^{-t}t^{a-1}dt.
 \end{equation}

\begin{figure}
\includegraphics[width=3in]{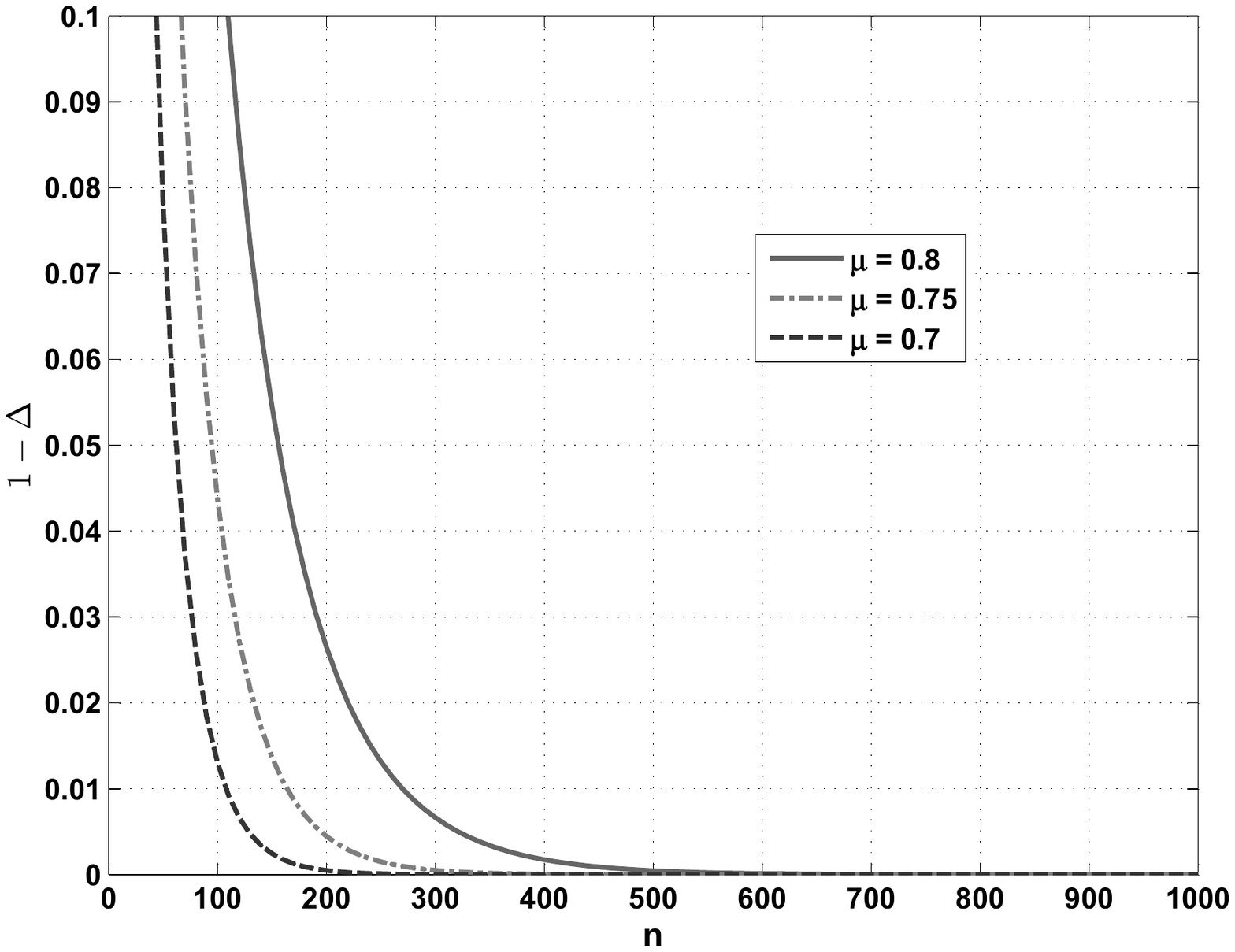}
\caption{$V_T(\bar{\mathbb{Q}}_X, \mathbb{Q}_X)$ decreases rapidly as $n$ increases for typical $\mu$.}\label{Fig00}
\end{figure}

 Note that $0 < \mu < 1$, $1 - \Delta \rightarrow 0$ as $n\rightarrow \infty$ and the rapidity depends only on $\mu$ and is quite significant due to sphere hardening effect, which is shown in Fig. \ref{Fig00}. Actually, on one hand, $\mu$ can't be small, as it will lead to the fact that the achievability bound is not tight; On the other hand, $\mu$ can not be close to $1$ with small blocklength $n$ since the effect of truncation should be controlled. Hence, the choice of $\mu$ depends on $n$ and the bias between security and coding rate. Usually, $\mu \in [0.7, 0.85]$ is sufficient for $n \geq 400$. In this case, $V_T(\bar{P}_{X^n}, P_{X^n})$ will be small for most applications. The effect of truncation is regarded to be negligible with $n \geq 600$. As $V_T(\bar{P}_{X^n}, P_{X^n})$ upper bounds the penalty from the fact that the codewords are drawn from truncated Gaussian distribution but not real Gaussian distribution, the codewords can be regarded as Gaussian codewords with proper $\mu$ and moderately large $n$ from the perspective of the output ends in a statistical sense. In addition, $\mathbb{P}_1$ and $\mathbb{P}_0$ are Gaussian distributed, hence TVD between $\mathbb{P}_1$ and $\mathbb{P}_0$ is strongly related to the power level $P(n)$. From Pinsker's inequality
\begin{equation}\label{UPP}
V_T(\mathbb{P}_1,\mathbb{P}_0)\leq \sqrt{\frac{1}{2} D(\mathbb{P}_1\|\mathbb{P}_0)},
\end{equation}
where $D(\mathbb{P}_1\|\mathbb{P}_0)$ is KL divergence between them. As the coordinates of $\mathbb{P}_0$ and $\mathbb{P}_1$ are both i.i.d normal distributed,  $D(\mathbb{P}_1\|\mathbb{P}_0)$ is convenient to evaluate due to the chain rule of KL divergence. The last inequality can be used to deduce an upper bound of $V_T(\mathbb{P}_1,\mathbb{P}_0)$. If $P(n)$ is small enough such that
\begin{equation}\label{UPP2}
\begin{split}
\delta = &1- \Delta + \sqrt{\frac{1}{2} D(\mathbb{P}_1\|\mathbb{P}_0)}\\
\geq & V_T(\bar{P}_{X^n}, P_{X^n}) + V_T(\mathbb{P}_1, \mathbb{P}_0)\\
\geq & V_T(\bar{\mathbb{P}}_1, \mathbb{P}_0),
 \end{split}
 \end{equation}
 the TVD constraint $V_T(\bar{\mathbb{P}}_1, \mathbb{P}_0) \leq \delta$ is surely satisfied.
If we directly let $\sqrt{\frac{1}{2} D(\mathbb{P}_1\|\mathbb{P}_0)} = \delta$, i.e., neglect the effect of truncation when determining the maximum transmission power $P(n)$, it will lead to a larger achievability bound and also a larger converse bound due to more optimistic estimations of $P(n)$.

By utilizing proper $\mu$ and $P(n)$, these bounds can be used to estimate the maximal throughput under covert constraint given by TVD.
  \subsection{Numerical Results on the Throughput}
  In this section, the previous results are applied to characterize the allowable throughput numerically under a given covert constraint and error probability of decoding.
  With given blocklength $n$ , error probability $\epsilon$ and covert constraint $V_T(\bar{\mathbb{P}}_1, \mathbb{P}_0) \leq \delta$, the computation process is described as follows,
  \begin{itemize}
  \item [(1):] Choose $0.7 \leq \mu \leq 0.85$, so that $1- \Delta < \delta$ is satisfied.
  \item [(2):] Solve the equation $D(\mathbb{P}_1\|\mathbb{P}_0) = 2(\delta + \Delta - 1)^2$ to get the value $P(n)$.
  \item [(3):]  Calculate the achievability bound from (\ref{AC3}) and the converse bound from (\ref{EandM}). The details of the calculation of the achievability bound are listed in Section \ref{Calculation}, and the calculation of the converse bound is nearly the same as that of \cite{Polyanskiy} except that we have a $P(n)$ for each $n$.
  \end{itemize}

   In the following figures,  we choose $\mu = 0.8$ and the least blocklength is $200$. (When $n =100$, the effect of truncation is notable and the covert constraint $V_T(\bar{\mathbb{P}}_1, \mathbb{P}_0) \leq \delta$ may be violated). With these choices of blocklength $n$, the effect of truncation is upper bounded by $1-\Delta$ from (\ref{trunction}) and is under $0.04$ from Fig.\ref{Fig00}. In fact, we can choose smaller $\mu$ (such as $\mu = 0.7$) so that the effect of truncation is almost under $0.005$. To solve the equation in the above step (2), we use
   \begin{equation}
    D(\mathbb{P}_1,\mathbb{P}_0) = \frac{n}{2}\left[P(n) - \ln (1 + P(n))\right]\log e,
   \end{equation}
   since we assume that the normal distribution of the background noise has variance $1$.

\begin{figure}
\includegraphics[width=3in]{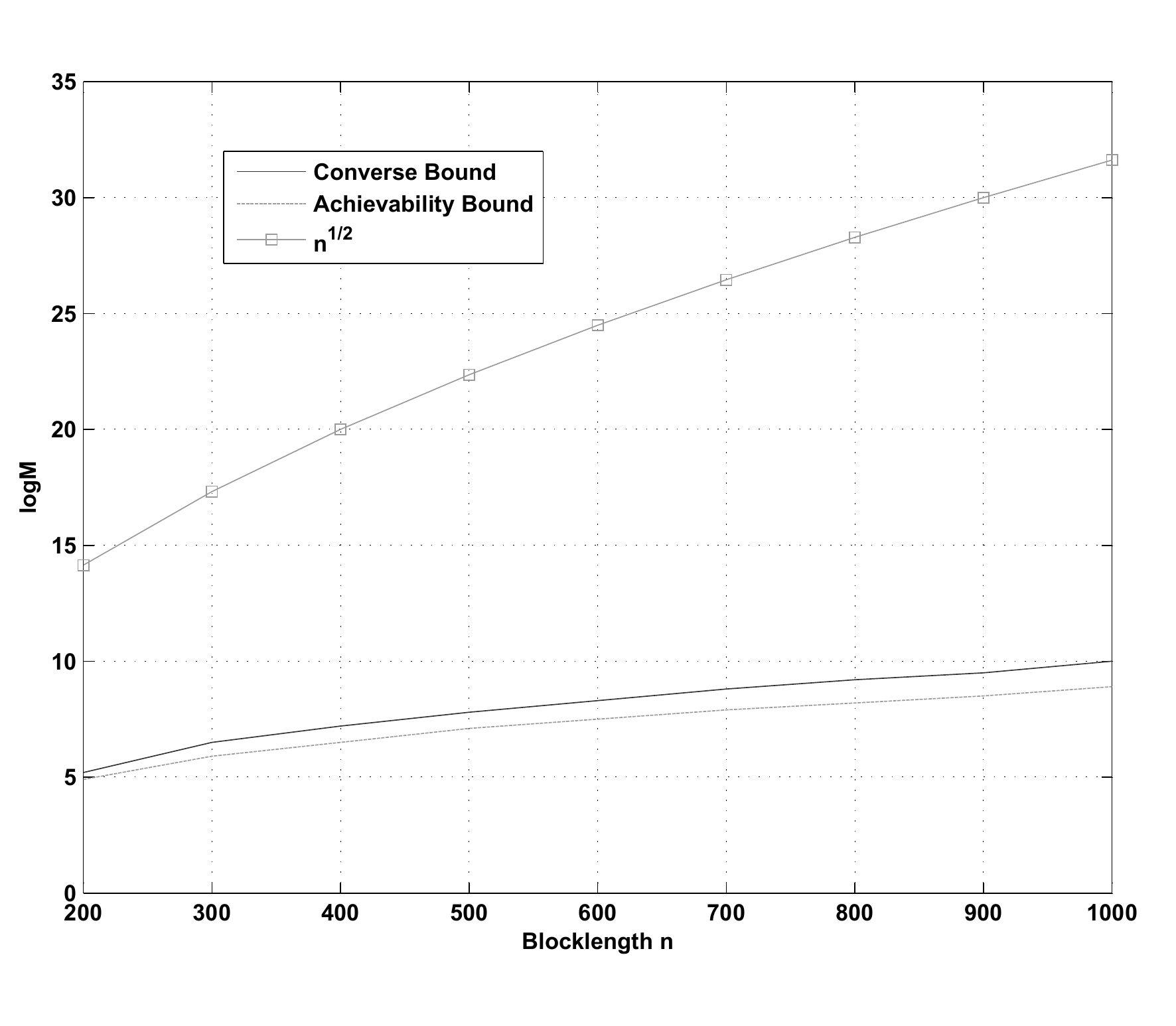}
\caption{The achievability and converse bounds where $\epsilon = 0.01$, $\delta = 0.1$ and $\mu = 0.8$.}\label{Fig6}
\end{figure}

\begin{figure}
\includegraphics[width=3in]{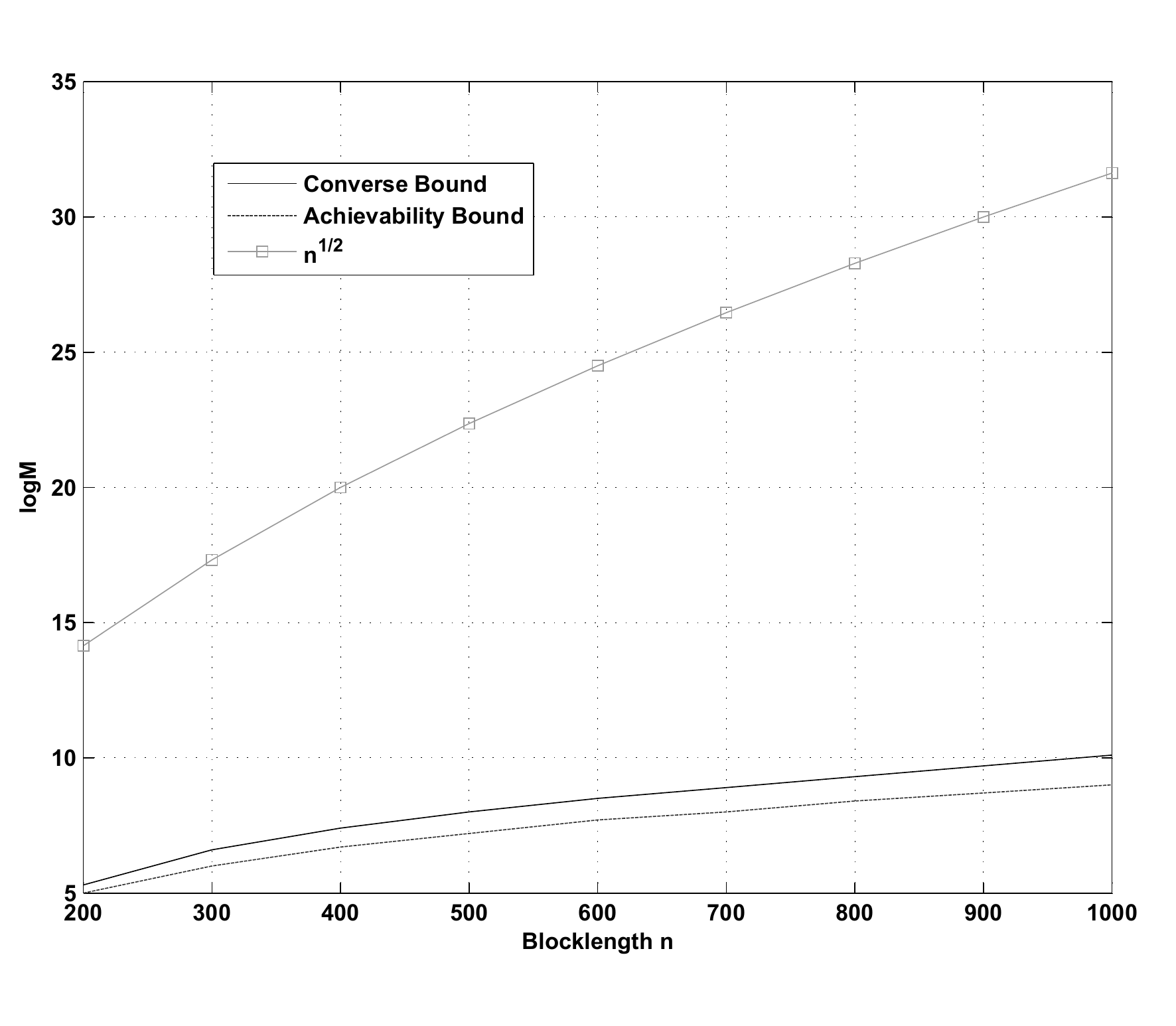}
\caption{The achievability and converse bounds where $\epsilon = 0.1$, $\delta = 0.1$ and $\mu = 0.8$.}\label{Fig7}
\end{figure}

\begin{figure}
\includegraphics[width=3in]{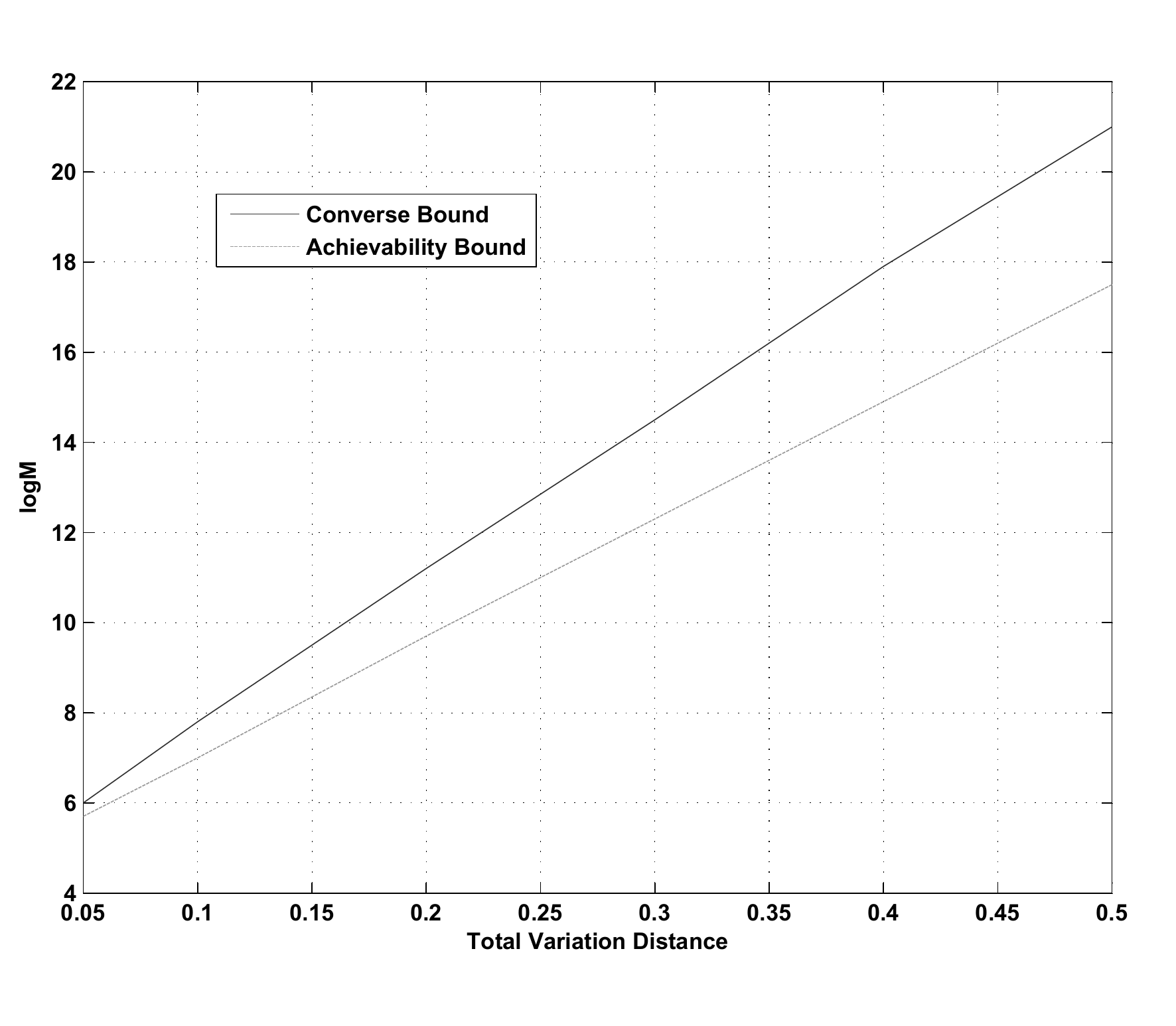}
\caption{The achievability and converse bounds with varying requirements of $V_T(\bar{\mathbb{P}}_1, \mathbb{P}_0)$, $\epsilon = 0.01$, $n = 500$ and $\mu = 0.8$.}\label{Fig8}
\end{figure}

\begin{figure}
\includegraphics[width=3in]{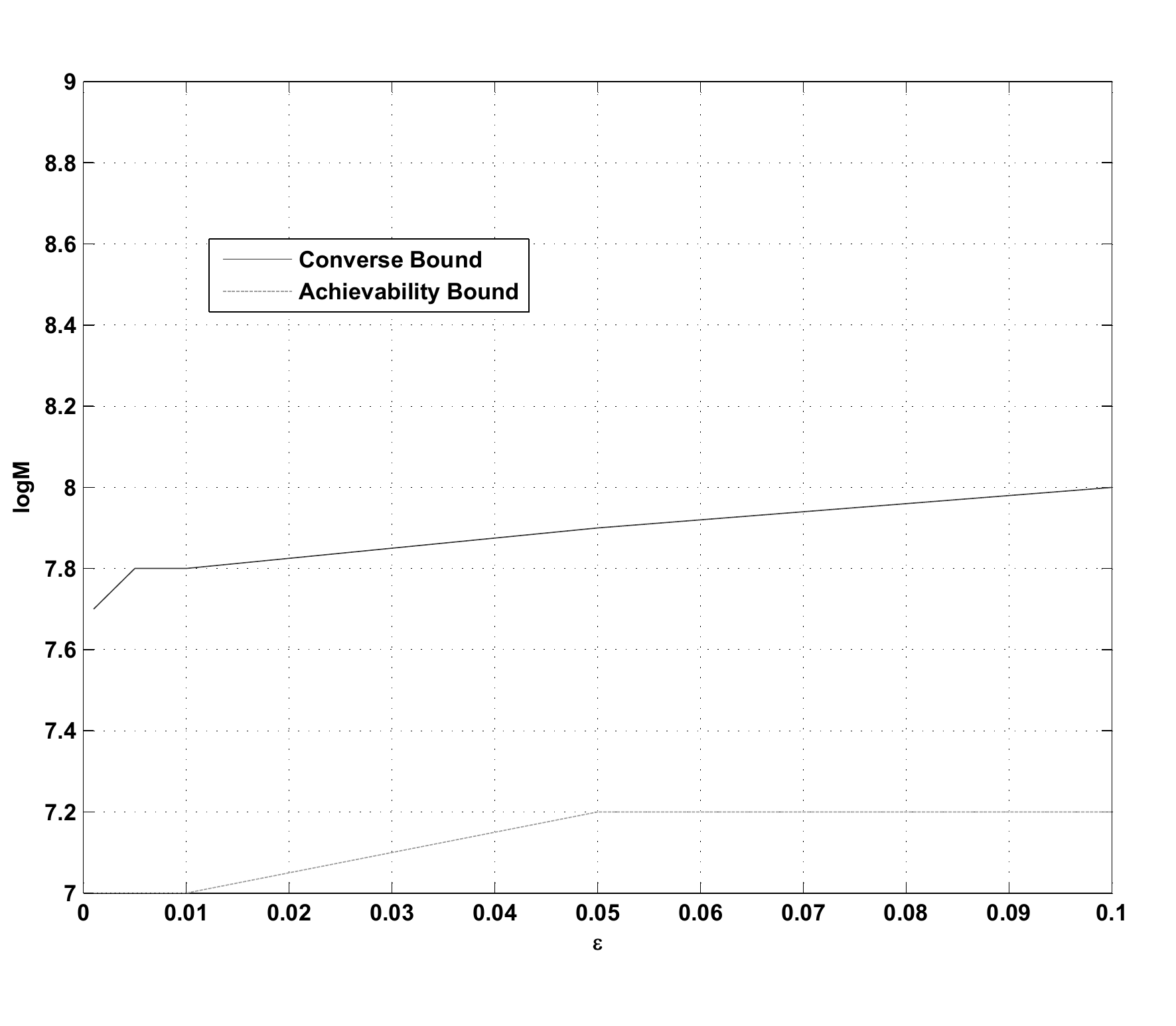}
\caption{The achievability and converse bounds with varying $\epsilon$,  $\delta = 0.1$, $n = 500$ and $\mu = 0.8$.}\label{Fig9}
\end{figure}

    In Fig.\ref{Fig6} and Fig.\ref{Fig7}, the bounds are plotted with varying values of blocklength. They show that the achievable throughput is much less than $\sqrt{n}$ when the upper bound of total variation distance is $\delta = 0.1$ or even less, which is quite surprising. We can explain it from several aspects. Firstly, the covert constraint $V_T(\bar{\mathbb{P}}_1, \mathbb{P}_0) \leq 0.1$ imposes severe limitation on the power level. Secondly, the utilization of inequalities makes the power level in the bound underestimated. Thirdly, the achievability bound is not the tightest, as we have explained in Section \ref{Calculation}. Fourthly, in contrast with conventional knowledge on communication theory, our results do not consider the effect of degree of freedom, such as bandwidth and time interval, which are necessary ingredients and will improve its practical utility in covert communication. In Fig.\ref{Fig8}, we plot the allowable throughput with fixed blocklength $n = 500$ but varying $\delta$. we can see that the throughput grows almost linearly with $\delta$ which is directly related with the power. There is a similar fact in asymptotic situation: the capacity grows linearly with the power in a channel with infinite bandwidth \cite{Cover}.  In Fig.\ref{Fig9}, we plot the allowable throughput under varying maximal probability of error $\epsilon$ with fixed blocklength $n= 500$. It is obvious that the effect of error probability is negligible. We will explain it as follows. From the well known result of \cite{Polyanskiy}, the error probability will affect the throughput by the term of second order asymptotics - the channel dispersion. However, as the power is so low that both the first and second order asymptotics are small, the variation of the error probability in the inverse $Q$ function will have little influence on the throughput. The upper and lower bounds on $\log_2(M)$ are quite close to each other, as shown in Fig.6, Fig.7, Fig.8 and Fig.9. This fact suggests that the tightness and consistency of our derived bounds under covert constraints imposed in terms of bounds on TVD distances. Therefore, the results provide accurate characterization for the attainable throughput of covert communication.


\section{Conclusion}
The paper investigated the finite blocklength performance of covert communication over AWGN channels when the covert constraint is in the form of total variation distance. Due to the limitations of previous results in finite blocklenth regime, general achievability and converse bounds on random coding are considered and then the bounds under Gaussian random coding are derived. We provide further discussions on the bounds and previous ones. It is shown that optimal distributions for the achievability bound and converse bound under random coding with maximal power constraint over AWGN channels are solutions of two minimax problems where total variation distance is the objective function. In the end, the bounds are applied in seeking limits of covert communications over AWGN channels.

\appendices
\section{Proof of Theorem 4}\label{Proof4}

\begin{Proof}
The proof is an extension on that of Theorem 65 in \cite{Polyanskiy} since $P(n)$ is a function of $n$. Here we should be careful about the order of the moments related to $P(n)$ in normal approximation.

Denote $A = \mathbb{R}^n$, $B = \mathbb{R}^n$ and $P_{Y^n|X^n = x^n} = \mathcal{N}(x^n, \mathbf{I}_n)$.
The information density $i(x^n; Y^n)$ under $P_{Y^n|X^n = x_0}$ with $x_0 = (\sqrt{P(n)}, \sqrt{P(n)}, \cdots, \sqrt{P(n)})$ can be expressed as
\begin{equation}\label{H_n}
\begin{split}
H_n = & n\log\sigma_Y + \frac{nP(n)}{2\sigma_Y^2}\log e \\
+ & \frac{1}{2\sigma_Y^2}\log e \sum_{i=1}^n ((1 - \sigma_Y^2) Z_i^2 + 2\sqrt{P(n)Z_i})\\
= & \frac{n}{2}\log(1+P(n))\\
 + & \frac{1}{2}\frac{P(n)}{(1 + P(n))}\sum_{i+1}^n (1 - Z_i^2 + \frac{2}{\sqrt{P(n)}}Z_i)\log e\\
= & nC(n) - \sum_{i=1}^n S_i(n)
\end{split}
\end{equation}
where $Z_i\sim \mathcal{N}(0,1), i= 1,\cdots, n$, $ C(n) = \frac{1}{2}\log (1+P(n))$,
and $S_i(P(n)) = \frac{P(n)\log e}{2(1 + P(n))}(Z_i^2 - 2\frac{Z_i}{\sqrt{P(n)}} - 1)$.
We have $E(S_i) = 0$ and
\begin{equation}\label{VN}
\begin{split}
V(n)& = V(P(n)) = Var(S_i)\\
 = &\left[\frac{P(n)\log e}{2(1 + P(n))}\right]^2\left(2 + \frac{4}{P(n)}\right).
\end{split}
\end{equation}

In the analysis of the asymptotical performance on the convergence of formula (\ref{H_n}) by the central-limit theorem,
we define $B(P(n)) = 6 \mathbb{E}(|S_i(P(n))^3|)/V(P(n))^{\frac{3}{2}}$, and
will claim that the quantity $B(P) = \frac{6E[|S_i(P)|^3]}{V(P)^{\frac{3}{2}}}$ is bounded as $P \rightarrow 0$.

From the fact that $S_i(P) = \frac{P\log e}{2(1 + P)}(Z_i^2 - 2\frac{Z_i}{\sqrt{P}} - 1)$
and $V(P) = \left[\frac{P\log e}{2(1 + P)}\right]^2\left(2 + \frac{4}{P}\right)$, we have
\begin{equation*}
\begin{split}
|S_i(P)|^3  = & [\frac{P\log e}{2(1 + P)}]^3 |Z_i^2 - 2\frac{Z_i}{\sqrt{P}} - 1|^3\\
\leq &[\frac{P\log e}{2(1 + P)}]^3|Z_i^2 + 2\frac{Z_i}{\sqrt{P}} + 1|^3\\
=  &[\frac{P\log e}{2(1 + P)}]^3 |Z_i^6 + \frac{8Z_i^3}{P^{\frac{3}{2}}} + 1
+ \frac{6Z_i^5}{\sqrt{P}}  \\
+ &3Z_i^4 + \frac{12Z_i^2}{P}(1 + Z_i^2)+ 3Z_i^2 + \frac{6Z_i}{\sqrt{P}} + \frac{12Z_i^3}{\sqrt{P}}|\\
\leq & [\frac{P\log e}{2(1 + P)}]^3 (Z_i^6 + \frac{8|Z_i|^3}{P^{\frac{3}{2}}} + 1 + \frac{6|Z_i|^5}{\sqrt{P}} + 3Z_i^4 \\
 + &\frac{12Z_i^2}{P}(1 + Z_i^2)+ 3Z_i^2 + \frac{6|Z_i|}{\sqrt{P}} + \frac{12|Z_i|^3}{\sqrt{P}}),\\
\end{split}
\end{equation*}
\begin{equation}\label{BoundTP}
\begin{split}
\mathbb{E}[|S_i(P)|^3] \leq & [\frac{P\log e}{2(1 + P)}]^3 (\mathbb{E}[Z_i^6] + \frac{8\mathbb{E}[|Z_i|^3]}{P^{\frac{3}{2}}} + 1\\
 +& \frac{6\mathbb{E}[|Z_i|^5]}{\sqrt{P}} + 3\mathbb{E}[Z_i^4] + \mathbb{E}[\frac{12Z_i^2}{P}(1 + Z_i^2)]\\
 + &3\mathbb{E}[Z_i^2] + \frac{6\mathbb{E}[|Z_i|]}{\sqrt{P}} + \frac{12\mathbb{E}[|Z_i|^3]}{\sqrt{P}})\\
= & [\frac{P\log e}{2(1 + P)}]^3 [C_1 + \frac{C_2}{P} + \frac{16\sqrt{2}}{\sqrt{\pi}}\frac{1}{P^{\frac{3}{2}}} ],
\end{split}
\end{equation}
where $C_1, C_2, C_3$ are positive constants. Hence, as $P(n)\rightarrow 0$,
\begin{equation}\label{BoundBP}
\begin{split}
0 < B(P(n)) < \frac{[\frac{P\log e}{2(1 + P)}]^3 [C_1 + \frac{C_2}{P} + \frac{16\sqrt{2}}{\sqrt{\pi}} \frac{1}{P^{\frac{3}{2}}} ]}{[\frac{P\log e}{2(1 + P)}]^3(2 + \frac{4}{P})^{\frac{3}{2}}} = O (1).
\end{split}
\end{equation}
Thus, the quantity $\frac{B(P)}{\sqrt{n}}$ approaches 0 as $n\rightarrow \infty$.

Then for the analysis of the coding rate, we consider the quantity
$\alpha(n) = 1-\epsilon - \frac{2B(P(n))}{\sqrt{n}}$
with fixed $\epsilon$ as an increasing function of $n$, which is positive if $2B(P(n)) < (1-\epsilon)\sqrt{n}$.
Denote $\zeta_n = \sqrt{nV(n)}Q^{-1}(\alpha_n)$.  As $S_i,i = 1,\cdots,n$ are i.i.d zero- mean variables with variance $V(P(n))$,
Berry Essen Theorem implies that
\begin{equation}
\mathbb{P}[\sum_{i=1}^n S_i \leq \zeta_n] \leq \alpha(n) + \frac{B(P(n))}{\sqrt{n}}.
\end{equation}
If we further let $\log \gamma = -\zeta_n + nC(n) = \sqrt{nV(n)}Q^{-1}(\alpha_n) + nC(n)$ and $\alpha = 1- \epsilon$,
from the inequality
\begin{equation}
\beta_{\alpha}(x,Q_Y) \geq \sup_{\gamma >0}\frac{1}{\gamma}(\alpha - P_{Y|X=x}[\frac{dP_{Y|X=x}}{dQ_Y} \geq \gamma]),
\end{equation}
we further have
\begin{equation}
\begin{split}
\beta_{1 - \epsilon}^n &\geq \sup_{\gamma >0}\frac{1}{\gamma}(1-\epsilon - \mathbb{P}[H_n \geq \log \gamma ])\\
& =   \sup_{\gamma >0}\frac{1}{\gamma}(1-\epsilon - \mathbb{P}[nC(n) - \sum_{i=1}^n S_i(n) \geq \log \gamma ]) \\
& = \sup_{\gamma >0}\frac{1}{\gamma}(1-\epsilon - \mathbb{P}[\sum_{i=1}^n S_i(n) \leq \zeta_n ])\\
& \geq e^{\zeta_n - nC(n)}\{\frac{2B(P(n))}{\sqrt{n}} + \alpha(n) - \mathbb{P}[\sum_{i=1}^n S_i(n) \leq \zeta_n ]\}\\
& = e^{\zeta_n -n C(n)} \frac{B(P(n))}{\sqrt{n}}.
\end{split}
\end{equation}
Since the codewords are under equal power constraint, from (\ref{Conv1}), we have
\begin{equation*}
\log M^*_e(n,\epsilon, P(n)) \leq nC(n) - \zeta_n + \frac{1}{2}\log n - \log B(P(n)).
\end{equation*}
In addition, since \begin{equation}
\begin{split}
\zeta_n  = &-\sqrt{nV(n)}Q^{-1}(\alpha_n)\\
 = &- \sqrt{nV(n)}Q^{-1}(1-\epsilon - \frac{2B(P(n))}{\sqrt{n}})
\end{split}
\end{equation}
for some $\theta \in [1-\epsilon - \frac{2B(P(n))}{\sqrt{n}}, 1-\epsilon]$, we get
\begin{equation*}
\begin{split}
\zeta_n  = &-\sqrt{nV(n)}Q^{-1}(\alpha_n) \\
= &- \sqrt{nV(n)}Q^{-1}(1 -\epsilon) + 2B(P(n))\sqrt{V}\frac{dQ^{-1}}{d x}(\theta)
\end{split}
\end{equation*}
 where the term $\frac{dQ^{-1}}{d x}(\theta)$ can also be lower bounded by $g_1(P(n), \epsilon) = \min_{\theta \in [1- 2\epsilon, 1-\epsilon]} \frac{dQ^{-1}}{d x}(\theta)$. Note that here we use the fact that $\frac{B(P(n))}{\sqrt{n}}$ approaches $0$ as $n \rightarrow \infty$. Thus,
\begin{equation}
\begin{split}
&\log M^*_e(n,\epsilon, P(n)) \leq  nC(n) + \sqrt{nV(n)}Q^{-1}(1 -\epsilon) \\
- &2B(P(n))\sqrt{V(n)}\frac{dQ^{-1}}{d x}(\theta) + \frac{1}{2}\log n - \log B(P(n)).
\end{split}
\end{equation}
Considering the order of each term, we have
\begin{equation}\label{Conv5}
\begin{split}
\log M^*_e(n, \epsilon, P(n))& \\
\leq nC(n) &-\sqrt{nV(n)})Q^{-1}(\epsilon) + \frac{1}{2}\log n + O(1).
\end{split}
\end{equation}
Consequently, the converse bound is proved when $P$ is a function of $n$.
\end{Proof}
\section{Proof of Theorem \ref{AC1}}\label{Proof6}
\begin{Proof}
The major difference of the proof from Theorem 67 is that the radius of the codewords varies. Thus the information density function is related to the radius, which brings more complexity.
Some definitions and notions used in the proof are listed as follows.
 \begin{enumerate}
 \item The codeword is now $x^n \in \mathsf{A} \triangleq \mathbb{R}^n$.
 \item $P_{Y^n|X^n=x^n} $ stands for the condition probability of $Y^n$ when the codeword $x^n$ is sent.
 \item Let $\mathsf{F}_n = \{x^n: \|x_i\|^2 \leq nP(n)\}$. It is clear that $\bar{\mathsf{F}}_n$ is a subset of $\mathsf{F}_n$.
\end{enumerate}

{\bfseries Generation of the codebook}: The process of generation is the same as in Lemma \ref{Le1} and the dependent test is substituted by Neyman- Pearson test as in Theorem \ref{AcTheorem1}. For each n, the distribution $P_X$ is $\mathcal{N}(0 ,\mu P(n)\bm{I}_n)$ where $0 <\mu <1$ will be determined later, i.e, each coordinate of these candidates is i.i.d drawn from $\mathcal{N}(0, \mu P(n))$. Each codeword is randomly chosen from the set $\bar{\mathsf{F}}_n$ following the steps in Lemma \ref{Le1}. The conclusion is an application of Corollary \ref{AcCorollary1}. The details are as follows.

Since the candidates of these codewords are generated from $\mathcal{N}(\bm{0} ,\mu P(n)\bm{I}_n)$, the auxiliary distribution $P_{Y^n} = \mathcal{N}(\bm{0}, (1+\mu P(n))\bm{I}_n)$
As our bounds are based on binary hypothesis test between $P_{Y^n|X^n=x^n}$ and $P_{Y^n}$, it is necessary to evaluate $\beta_{\alpha}^n(x^n, \mathcal{N}(\bm{0}, (1+\mu P(n))\bm{I}_n))$ with a given detection probability $\alpha$. Assume $x^n_R=[\sqrt{R(n)}, \cdots, \sqrt{R(n)}]$,  because spherical symmetry will lead to the same $\beta$ with given $\alpha $ on the surface with radius $\sqrt{nR}$.
Under $P_{Y^n}$ and $P_{Y^n|X^n =x^n_R}$, the expressions of $\beta_{\alpha}^n$ and $\alpha = 1 + \tau(n) -\epsilon$ are
\begin{equation}
\beta_{1-\epsilon + \tau(n,R)} = \mathbb{P}[G_n(R) \geq \gamma(n,R)]
\end{equation}
with
\begin{equation}\label{GnR}
\begin{split}
G_n(R) = & \frac{n}{2}\log(1+\mu P(n)) -\frac{nR(n)}{2}\log e \\
+ &\frac{1}{2}\log e \sum_{i=1}^n\left(2\sqrt{R(n)(1+ \mu P(n))}Z_i - \mu P(n) Z_i^2 \right)
\end{split}
\end{equation}
and
\begin{equation}
\alpha = 1- \epsilon + \tau(n,R) = \mathbb{P}[H_n(R) \geq \gamma(n,R) ]
\end{equation}
with
\begin{equation}\label{HnR}
\begin{split}
H_n(R) = &\frac{n}{2}\log(1 + \mu P(n)) + \frac{nR(n)}{2(1+\mu P(n))}\log e \\
+ &\frac{\log e}{2(1+\mu P(n))}\sum_{i=1}^n\left(2\sqrt{R(n)}Z_i -\mu P(n)Z_i^2\right).
\end{split}
\end{equation}
Note that $R = R(n) \in [0, P(n)]$ and the parameter $\gamma(n, R)$ is determined by the detection probability $1- \epsilon + \tau(n,R)$, hence is determined by $\tau(n,R)$ when $\epsilon$ is given.
First, we rewrite $H_n$ as
\begin{equation}
\begin{split}
H_n = &\frac{n}{2}\log(1 + \mu P(n)) + \frac{nR}{2(1+\mu P(n))}\log e \\
+ &\frac{\log e}{2(1+\mu P(n))}\sum_{i=1}^n\left(2\sqrt{R}Z_i -\mu P(n)Z_i^2\right)\\
= & nC_{\mu}(n) + \frac{nR\log e}{2(1+\mu P)} - \sum_{i=1}^n S_i
\end{split}
\end{equation}
where $S_i = \frac{\log e}{2(1+\mu P)} \left[\mu PZ_i^2 - 2\sqrt{R}Z_i\right]$.
It is easy to get
$E[S_i] = \frac{\mu P\log e}{2(1+\mu P)}$  and
\begin{equation}
\begin{split}
\hat{V}(P) = Var(S_i) = &\left(\frac{\log e}{2(1+\mu P)}\right)^2(4R+ 2\mu^2 P^2) \\
= &V_{\mu} \cdot \left(\frac{2R+ \mu^2 P^2}{2\mu P+ \mu^2P^2}\right)
\end{split}
\end{equation}
where $V$ is the channel dispersion with power $\mu\cdot P$ and further denote
\begin{equation}
\begin{split}
T_{\mu}(P) = &\mathbb{E}\left[|S_i - E(S_i)|^3\right]\\
= & \mathbb{E}\left[|\frac{\log e}{2(1+\mu P)}[\mu P + 2\sqrt{R}Z_i -\mu PZ_i^2]|^3\right]
\end{split}
\end{equation}
and
$B_{\mu}(P) = \frac{6T_{\mu}(P)}{\hat{V}_{\mu}(P)^{3/2}}$.

Denote $\hat{S}_i = S_i - E(S_i)$, then
\begin{equation}
H_n = nC_{\mu}(n) + \frac{nR\log e}{2(1+\mu P)} - \frac{n\mu P\log e}{2(1+\mu P)} - \sum_{i=1}^n \hat{S}_i.
\end{equation}
\noindent Let
 $\alpha^{\mu}_n = 1 -\epsilon + \tau^{\mu}_n(R)$ be substitute for $\alpha = 1 + \tau(n) -\epsilon$, $\zeta_n^{\mu} = \sqrt{n\hat{V}_{\mu}(n)}Q^{-1}(\alpha_n^{\mu})$
  and
  \begin{equation}
  \begin{split}
  \log \gamma_n = &nC_{\mu}(P(n)) + \frac{n(R-\mu P)\log e}{2(1+\mu P)} \\
  + &\sqrt{n\hat{V}_{\mu}(P(n))}Q^{-1}(\alpha_n^{\mu}).
  \end{split}
  \end{equation}

\noindent Similar as (\ref{BoundTP}) and (\ref{BoundBP}), $\frac{B_{\mu}(P)}{\sqrt{n}}$ tends to $0$ as $n\rightarrow \infty$. $\alpha^{\mu}_n$ is certainly less than $1$ when $n$ is sufficiently large and the definition of $\zeta^{\mu}_n$ is meaningful.
As $\hat{S}_i,i = 1,\cdots,n$ are i.i.d zero- mean variables with variance $\hat{V}(P(n))$, Berry-Esseen Theorem implies that
\begin{equation}
\mathbb{P}[\sum_{i=1}^n \hat{S}_i \leq \zeta^{\mu}_n] \geq \alpha^{\mu}_n - \frac{B_{\mu}(P(n))}{\sqrt{n}}
\end{equation}
\noindent for all codewords $x^n$ with the same radius $\sqrt{nR(n)}$ in the space.
\newcounter{TempEqCnt9}
\setcounter{TempEqCnt9}{\value{equation}}
\setcounter{equation}{92}
\begin{figure*}[!t]
\normalsize
\begin{equation}\label{Uppbeta}
\begin{split}
\log \beta^n_{1-\epsilon + \tau_n}(x_R^n, Q_Y) \leq& \log \beta_{\alpha_n}\\
 = &\mathbb{E}[exp{-i(x^n; Y^n)}1_{i(x^n;Y^n) \geq \log \gamma_n}|X^n = x^n]\\
\leq & \log [\frac{1}{\sqrt{n}\gamma_n}(\frac{2\log 2}{\sqrt{2\pi \hat{V}(n)}}+ 4B_{\mu}(P(n)))]\\
= & -\frac{1}{2}\log n - nC_{\mu}(P(n)) - \frac{n(R-\mu P)\log e}{2(1+\mu P)} - \sqrt{n\hat{V}_{\mu}(P(n))}Q^{-1}(\alpha_n) + \log \left(\frac{2\log 2}{\sqrt{2\pi \hat{V}_{\mu}(n)}}+ 4B_{\mu}(P(n))\right)\\
= &  - nC_{\mu}(P(n)) - \frac{n(R-\mu P)\log e}{2(1+\mu P)} - \sqrt{nV_{\mu}(P(n))}Q^{-1}(\alpha_n)\cdot \left(\frac{2R+ \mu^2 P^2}{2\mu P+ \mu^2P^2}\right)^{\frac{1}{2}}\\
+ &\log \left[\frac{2\log 2}{\sqrt{2\pi V(n)}}\cdot \left(\frac{2R+ \mu^2 P^2}{2\mu P+ \mu^2P^2}\right)^{\frac{1}{2}}+ 4B_{\mu}(P(n))\right] -\frac{1}{2}\log n.
\end{split}
\end{equation}
\hrulefill
\vspace*{4pt}
\end{figure*}
\setcounter{equation}{93}%

From Lemma 47 in \cite{Polyanskiy}, an upper bound (\ref{Uppbeta}) of $\beta^n_{1-\epsilon + \tau_n}(x_R^n, Q_Y)$ for $x_R^n \in \mathsf{F}_n$ is found for each $n$.

From now on, the blocklength $n$ is sufficiently large and fixed so that $\alpha^{\mu}_n = 1 - \epsilon + 2\tau^{\mu}_n(R)$ is less than 1. Note that $R$ varies in $[0, P(n)]$, to utilize Corollary \ref{AcCorollary1}, the following statements are important.
\begin{enumerate}

\item [(a)] From (\ref{BoundBP}), $B_{\mu}(P,R)$ is positive and bounded, $\tau^{\mu}_n(R) = \frac{B_{\mu}(P,R)}{\sqrt{n}}$ will be sufficiently small when $n$ is large. We can always find some $\tau_0$ so that the set $\bar{\mathsf{F}}_n$  is nonempty, and there are sufficiently many points in $\bar{\mathsf{F}}_n$.
\item [(b)] Since $P$ is a function of $n$, $\tau^{\mu}_n(R)$ is a continuous function of $R$ with fixed $n$. Considering the mapping from $[0, P(n)]$ of $\tau^{\mu}_n(R)$ to $\mathbb{R}$, it is proper since $[0, P(n)]$ is compact and $\tau^{\mu}_n(R)$ is continuous.\footnote{ A continuous map: $f: X \rightarrow Y$ between topological spaces is called proper if for every compact subspace $K \subseteq Y$, the pre-image $f^{-1}(K)$ is compact. When $X$ is compact and $Y$ is Hausdorff, then every continuous map $f: X \rightarrow Y$ is proper.} When the range of $\tau^{\mu}_n(R)$ is constrained to be $[\tau_0, \frac{n}{n+1}\epsilon]$, the preimage $R$ of $\tau^{\mu}_n(R)$ is a compact set. We denote it as $\mathsf{R}_n$.
\item [(c)]From the expression (\ref{maxim}), it is a continuous function of $R$. Consequently, we can always find some $R^*\in \mathsf{R}_n$ which maximizes (\ref{maxim}).
\item [(d)]As the set $\bar{\mathsf{F}}_n$ is determined by $\tau_0$ and $\sqrt{nR^*}$ is radius of the point $x_{R^*}^n = [\sqrt{R^*}, \cdots, \sqrt{R^*}]$ in $\bar{\mathsf{F}}_n$. The value $R^*$ depends on $\tau_0$.
\item [(e)]There are many choices of $\tau_0$, and we choose one which maximizes (\ref{logM1}).
The choice will lead to the tightest achievability bound for the throughput.
\end{enumerate}
Consequently, we have proved that a codebook exists which satisfies maximal power constraint and the lower bound of the size satisfies (\ref{Achi3}).
\end{Proof}

\end{document}